\documentclass[12pt]{iopart}

\usepackage{graphicx,color}
\usepackage{epsfig}
\usepackage[squaren]{SIunits}
\usepackage{textcomp}
\usepackage{nicefrac}

\begin{document}

\title[Condensation of Excitons in Cu$_2$O]{Condensation of Excitons in Cu$_2$O at Ultracold Temperatures: Experiment and Theory}
\author{Heinrich Stolz, Rico Schwartz, Frank Kieseling, Sunipa Som, Maria Kaupsch, Siegfried Sobkowiak, and Dirk Semkat}
\address{Institut f\"ur Physik, Universit\"at Rostock, D-18051 Rostock, Germany}
\ead{heinrich.stolz@uni-rostock.de}

\author{Nobuko Naka}
\address{Department of Physics, Kyoto University, Kyoto 606-8502, Japan}
\address{PRESTO, JST, 4-1-8 Honcho Kawaguchi, Saitama 332-0012, Japan}

%\author{Jan Brandt, Dietmar Fr\"ohlich, and Manfred Bayer}
%\address{Fakult\"at Physik, Technische Universit\"at Dortmund, D-44221 Dortmund, Germany}

\author{Thomas Koch and Holger Fehske}
\address{Institut f\"ur Physik, Ernst-Moritz-Arndt-Universit\"at Greifswald, D-17487 Greifswald, Germany}

\renewcommand{\baselinestretch}{1.0}

\newcommand{\be}{\begin{equation}}
\newcommand{\ee}{\end{equation}}
\newcommand{\bea}{\begin{eqnarray}}
\newcommand{\eea}{\end{eqnarray}}
\newcommand{\eps}{\varepsilon}
\newcommand{\ev}[1]{\langle {#1} \rangle}
\newcommand{\vc}[1]{\boldmath #1}
\newcommand{\ket[1]}{|{#1}\rangle}
\newcommand{\red[1]}{\textcolor{red}{#1}} 
\newcommand{\me}[0]{\mathrm{e}}
\newcommand{\mi}[0]{\mathrm{i}}
\newcommand{\md}[0]{\mathrm{d}}
%\newcommand{\micro}{\textmu}
%%%%%%%%%%%%%%%%%%%%%%%%%%%%%%%%%%%%%%%%%%%%%%%%%%%%%%

\begin{abstract}
We present experiments on the luminescence of excitons confined in a potential trap at milli-Kelvin bath temperatures under continuous-wave (cw) excitation. They reveal several distinct features like a kink in the dependence of the total integrated luminescence intensity on excitation laser power and a bimodal distribution of the spatially resolved luminescence. Furthermore, we discuss the present state of the theoretical description of Bose-Einstein condensation of excitons with respect to signatures of a condensate in the luminescence. 
The comparison of the experimental data with theoretical results with respect to the spatially resolved as well as the integrated luminescence intensity shows the necessity of taking into account a Bose-Einstein condensed excitonic phase in order to understand the behaviour of the trapped excitons.
\end{abstract}


\maketitle

\section{Introduction}

Almost 50 years ago, excitons \cite{blatt1962,moskalenko1962} have been suggested as particularly interesting candidates for Bose-Einstein %%@
condensation (BEC), as they consist of an electron and a hole in a semiconductor, both fermions bound to form a bosonic excitation and thus resembling %%@
most closely neutral atoms of usual matter. Due to their rather small mass comparable to the free electron mass, it was speculated that for exciton densities %%@
of the order of $10^{18}$\,cm$^{-3}$ -- easily achievable by absorption of photons -- critical temperatures of some 10\,K may be reached.

Due to their unique properties, the excitons of the so-called yellow series in the semiconductor cuprous oxide (Cu$_2$O) are still considered the most %%@
promising candidates for excitonic BEC \cite{froehlich1979,snoke2002sc,alvermann2011}. This is related to the large binding energy of 150\,meV, which shifts the Mott density to %%@
$3\cdot10^{18}$\,cm$^{-3}$ at cryogenic temperatures \cite{semkat2009,manzke2010}. Made up from doubly degenerate valence and conduction bands, the ground %%@
state of this series splits into the triply degenerate orthoexciton and the nondegenerate paraexciton, which is the energetically lowest exciton %%@
state, lying $\Delta=12.1$\,meV below the orthoexciton states. Due to the positive parity of the bands, the orthoexciton is only weakly optically allowed %%@
(quadrupole transition with oscillator strength $3\cdot 10^{-9}$ \cite{froehlich1991}), while the paraexciton as a pure triplet state with respect to the electron and hole spins \cite{waters} is optically %%@
forbidden in all orders. Its decay is only possible via an odd parity optical phonon resulting in a long lifetime in the microseconds range during %%@
which thermodynamic quasi-equilibrium may be reached. 

As in all physical systems for which BEC has been demonstrated up to now, excitons should be confined in a potential trap. This has the advantage that %%@
(i) the diffusion process, which reduces the exciton density, is suppressed and (ii) the critical number of particles required for the phase transition decreases much faster with %%@
temperature than in free space. The critical particle number is given by
\be\label{eq:Ncid}
N_{\rm crit}= \zeta(3) \left ( \frac{k_{\rm B} T}{\hbar \Omega_{\rm 0}} \right)^3\, ,
\ee
where $\Omega_{\rm 0}$ is the average oscillator frequency of the trapping potential and $\zeta$ denotes the Riemann Zeta function \cite{pethik2002}. 

Despite the promising properties mentioned above, all previous studies to create a dense gas of excitons in Cu$_2$O either in a bulk crystal or in a %%@
potential trap did not demonstrate conclusively excitonic BEC \cite{snoke2002sc,wolfe1986,snoke1990a,wolfe1993,snoke2000,gonokami2011}. 
The main reason for this failure turned out to be the existence of a very efficient exciton-exciton annihilation process that sets in at high exciton %%@
densities and whose rate scales with the square of the exciton density $n$ 
\be
\frac{dn}{dt}= - a n^2\quad .
\ee
Due to the rather large value of $a$ of the order of $10^{-16}\,{\rm cm}^3{\rm ns}^{-1}$, this process was believed to prevent the formation of a BEC of paraexcitons. 

However, in recent experiments using pulsed excitation, we have found in agreement with earlier studies \cite{snoke2002} that this process can be damped in shallow potential traps by almost two orders of magnitude \cite{schwartz2011}. While this allowed the accumulation of large exciton numbers, the temperatures in these experiments were still too high to undercut the critical temperature for Bose-Einstein condensation, which was predicted to require a bath temperature as low as 100 mK at the experimental conditions. Furthermore, the high pulse energies needed to create sufficient exciton numbers resulted in a substantial heating of the exciton gas during the laser pulse. Our strategy to overcome these problems was twofold: firstly, we reduced the temperature of the He bath as much as possible by preventing thermal radiation from the surrounding to reach the sample. In this way we reached a minimum temperature of 35 mK for zero incident laser power, which increases at laser powers of about 1 mW to 250 mK within one hour. Secondly, we switched over to cw-excitation, for which the possibility of creating large paraexciton numbers has been demonstrated recently \cite{sandfort2011}. 
By reducing the thermal load, we were able to get down to effective exciton temperatures as low as 200 mK at low excitation power. The main advantage of cw-excitation, however, is the possibility to achieve a quasi-equilibrium situation in which the decay rate of excitons is cancelled by the formation and relaxation rate of the species in the trap. This allows to drive the system through the possibly existing various phases by simply changing the excitation power of the driving laser, as was exemplified by the condensation experiments of exciton polaritons in semiconductor microcavities (for a review, see \cite{deng2010}). In contrast to polariton systems, the lifetime broadening for paraexcitons is extremely small due to their long lifetime of the order of $\micro$s. Therefore, we can neglect any damping effects due to the excitonic decay in the spectra \cite{littlewood,bayer}.

The paper is organised as follows: In section \ref{sec:exp}, we briefly sketch the experimental setup and present a typical set of experimental results which reflect a variety of experimental conditions obtained by changing the bath temperature and the excitation conditions. We concentrate in particular on the spatial profiles of the luminescence intensity and on the totally integrated intensity in dependence on the excitation power.
These results are analysed theoretically in section \ref{theory-1} calculating the excitonic luminescence of an interacting Bose gas on a mean field level. Thereby, we start with the usual assumption of global equilibrium and then extend the theory to the case of excitons in local thermodynamic equilibrium. Finally, we compare the theoretical results with the measurements and show that there is excellent qualitative agreement if we take into account the occurrence of a Bose-Einstein condensate of excitons. Section \ref{sec:concl} gives our conclusion and an outlook to further experiments.

%%%%%%%%%%%%%%%%%%%%%%%%%%%%%%%%%%%%%%%%%%%%%%%%%%%%%%%%%%%%%%%%%%%%%%%%%%%%%%%%%%%%%%%%%%%%%%%%%%%%%%

\section{Experiment}\label{sec:exp}
\subsection{Experimental setup}\label{sec:expsetup}

\begin{figure}[h]
 \begin{center}
 \begin{minipage}{14cm}
  \includegraphics[width=\textwidth]{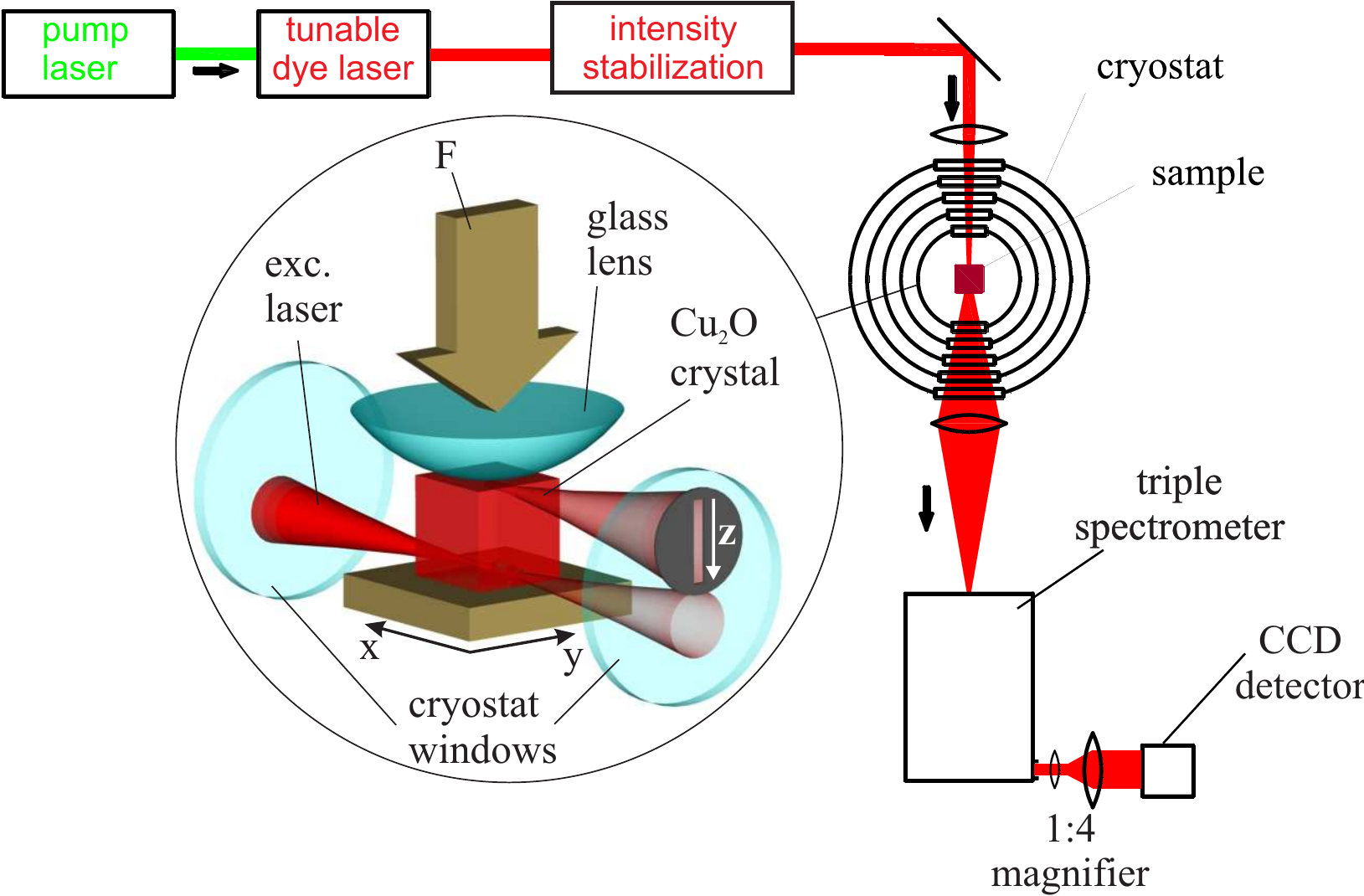}
 \end{minipage}
 \caption{Main features of the experimental setup. Excitons are created by a dye laser propagating along the $x$-direction normal to the strain axis ($z$-direction). The laser was focused either directly into the trap or onto a spot close to the trap. In the latter case the excitons diffuse towards the trap due to the gradient force initiated by the potential trap. The emission out of the trap is monitored spatially resolved along 
the $y$- and $z$-direction, while integrating along the $x$-direction.}
 \label{fig:aufbau}
 \end{center}
\end{figure}

For the studies at subkelvin temperatures, we used the same experimental setup as reported previously \cite{schwartz2011}, but implemented a narrow-band tunable dye laser (Coherent CR599, laser dye Rhodamin 6G) pumped by a 5W green solid state laser (Verdi 5), see figure \ref{fig:aufbau}. The laser power was stabilised by a closed feedback loop to within 1\%, the laser frequency and line width ($<0.5$~GHz) were measured with a wavemeter (High Finesse WS7, resolution 60 MHz). In order to enhance both the spectral and spatial resolution, we employed a fourfold magnification optical imaging system between the spectrometer exit slit and the detector.   

%%%%For the experiments at temperatures above 1.25 K the setup is described in detail in Ref. \cite{sandfort2011}.

For the experiments we used natural cuprous oxide crystals originally found in Namibia in the form of millimetre sized cubic specimens with well defined facets (see \cite{sandfort2011} for details). The quality of these samples was checked according to a low defect density, leading to %%@
long paraexciton lifetimes up to $1\,\rm{\micro s}$. For such samples, previous high resolution absorption measurements in a magnetic field revealed a %%@
paraexciton line width as narrow as $80\,\rm{neV}$, demonstrating their extremely high quality \cite{brandt2007}.

The potential trap for the confinement of the exciton gas was made by the well-known Hertzian stress technique %%@
\cite{wolfe1986,snoke2000,wolfe_ge,naka2002}, where a spherical stressor made of glass (radius 7.75\,mm) is pressed with a force $F$ against a flat %%@
surface of the crystal along a direction which we denote as $z$-direction (figure \ref{fig:aufbau}). As a result, a confining potential is generated in which the %%@
energies of ortho- and paraexcitons are lowered compared to the bulk. This potential $V_{\rm ext}(x,y,z)$ can be calculated from the known strain parameters of the yellow exciton states (for a recent %%@
calculation with refined parameters see \cite{sandfort2011}). To achieve agreement between the calculated potential profiles and the experimentally measured low energy border lines of the spatio-spectral %%@
images, we had to use as stressor radius a value 50 \% larger than the nominal one \cite{schwartz2011}. For a simple description, we decompose the potential trap and give in table \ref{parameter_ratenmodell} the parameters of the harmonic oscillator potential along $z$,  $V(z)=\alpha_{\|}(z-z_{\rm 0})^2-V_{\rm 0}$, and of the two-dimensional %%@
harmonic oscillator normal to $z$ in the $x,y-$plane, $V(x,y)=\alpha_{\perp} (x^2+y^2)$.
%%%%%%%%%%%%%%%%%%%%%%
\begin{table}
\begin{indented}
\item[]\begin{tabular}{ll} \hline\hline
parameter & value\\ \hline
potential curvature $\alpha_{\parallel}$ & 0.1334 $\micro$eV/$\micro$m$^2$\\
potential curvature $\alpha_{\perp}$ & 0.0733 $\micro$eV/$\micro$m$^2$\\
lifetime of paraexcitons & 650 ns \\
ortho-para conversion rate & 0.2 ns$^{-1}$ ${\dagger}$ \\
two-body decay rate $A_{\rm PP}$ & $2\cdot 10^{-18}$ cm$^{3}$ ns$^{-1}$ ${\ddagger}$ \\
two-body decay rate $A_{\rm OO}$ & $4.9\cdot 10^{-17}$  cm$^{3}$ ns$^{-1}$ \\
relaxation rate $\Gamma_{\rm rel}$ & $6\cdot 10^7$ s$^{-1}$ \\
potential trap minimum $V_{\rm 0}$ & 1.35 meV \\ \hline\hline \\
\end{tabular}
\caption[parameter_ratenmodell]{Parameters for trapping potential and rate model.\\
${\dagger}$ taken from reference \cite{snoke2002}\\
${\ddagger}$ due to the lower strain, the two-body decay rate $A_{\rm PP}$ is reduced by a factor of two compared to \cite{schwartz2011}}
\label{parameter_ratenmodell}
\end{indented}
\end{table}
%%%%%%%%%%%%%%%%%%%%%%%%

As resonant creation process of the excitons we used the indirect absorption process involving an odd parity optical $\Gamma _3^-$-phonon into the orthoexcitons in the trap (see inset in figure \ref{fig:rate_model}). These orthoexcitons quickly transform into paraexcitons. The laser beam was positioned about %%@
100\,$\micro$m away from the trap centre in positive $z$-direction, i.e., away from the stressor lens. In order to confine the primarily created orthoexcitons in the trap and to avoid any excitation outside the trap, which may lead to losses of the excitons, we tuned the energy of the %%@
laser photons slightly ($\approx 0.5$\,meV) below the onset of the phonon sideband in the bulk (at 2048.56\,meV).
By this process, about half of the incoming photons are transformed into excitons.  

To detect the primarily excited orthoexcitons for the calibration of the number of excited excitons, we have chosen the $\mathbf{k}$-vector of the excitation beam along the direction of observation. 

In the experimental setup (figure \ref{fig:aufbau}), the emitted light was imaged onto the entrance slit of a high resolution triple spectrograph (T64000, %%@
Jobin Yvon) usable either in subtractive or additive mode with a diffraction limited spatial resolution of the order of 10\,\micro m. The astigmatism %%@
of the spectrograph was corrected by a cylindrical lens (focal length $F=-1000$\,mm/ $F=1000$\,mm for subtractive and additive dispersion, %%@
respectively) in front of the entrance slit \cite{sandfort2011}. To obtain a $z$-resolved spectrum $I(z,\omega)$, along the direction of the applied %%@
strain, the luminescence from a small stripe of width $2\Delta y$ centred in the trap was integrated along the $x$-direction perpendicular to $z$ %%@
(see figure \ref{fig:aufbau}). Detection was done either by an intensified charge-coupled device (CCD) camera (Andor iStar) which could be gated with a minimum temporal resolution of 5 ns or with a nitrogen cooled CCD camera with high quantum efficiency (Andor Newton), which allowed long integration times. %%@

\subsection{Application of the rate model to cw-excitation}\label{sec:ratemodel}

With some adaptions, the rate model developed in \cite{schwartz2011} can also be applied to describe the cw experiments. These concern the following:
\begin{itemize}
\item 
Even under resonant excitation of excitons, a considerable density of unbound electron-hole pairs is generated \cite{naka2012}. This implies that the formation of an exciton from the hot electron-hole pair generated in the two-body decay (Auger-like process) takes a finite time comparable with other exciton relaxation times, e.g., the one related to phonon scattering. In a fast initial relaxation stage with a duration of several picoseconds \cite{kavoulakis1996a}, the carriers thermalize by longitudinal optical (LO) and longitudinal acoustic (LA) phonon scattering. For a strained crystal with an exciton potential trap, these electron-hole pairs will then undergo different relaxation scenarios depending on whether the strain causes a trapping  potential also for the unbound electron-hole pairs or not. In the latter case, the electrons and holes will diffuse from the place of generation (the potential trap) into the whole crystal and form excitons which themselves will drift again into the trap. In the former case, the electron-hole pairs will stay inside their trap and form a stable electron-hole plasma cloud with a density which is three orders of magnitude smaller than the exciton concentration according to the results in \cite{naka2012}. As derived in \ref{app:strain_eh}, strain leads to a trapping potential for unbound electron-hole pairs which is similar to that for the paraexcitons. To take this effect into account, we included in the rate model the unbound electron-hole pairs with a total number of $N_{\rm ehp}$. They are generated via the two-body decay of the ortho- and paraexcitons and recombine with a rate $\Gamma_{\rm rc} N_{\rm ehp}^2$. The recombination rate should depend on temperature in the same way as the two-body decay of the excitons ($\propto T^{-3/2}$). 
\item Due to heating and incomplete relaxation, the excitons may not cool down to bath temperature. We can describe this effect by assuming an effective exciton temperature $T_{\rm eff}$, a cooling time of 200 ns \cite{schwartz2011} and heating processes due to the energy release by ortho-para conversion $C_{\rm O}=10$~meV per exciton, Auger-like two-body decay with $C_{\rm XX}=2$~eV per exciton pair, and a non-radiative decay of paraexcitons (guessed release $C_{\rm P}=0.25$~eV per exciton). The temperature rise due to this heating is characterized by a constant $C_{\rm heat}$ which was adjusted to $C_{\rm heat}=7\cdot 10^{9}$~K/J.
\end{itemize}   

\begin{figure}[b]
 \begin{center}
% \begin{minipage}{14cm}
  \includegraphics[width=0.7\textwidth,angle=-90]{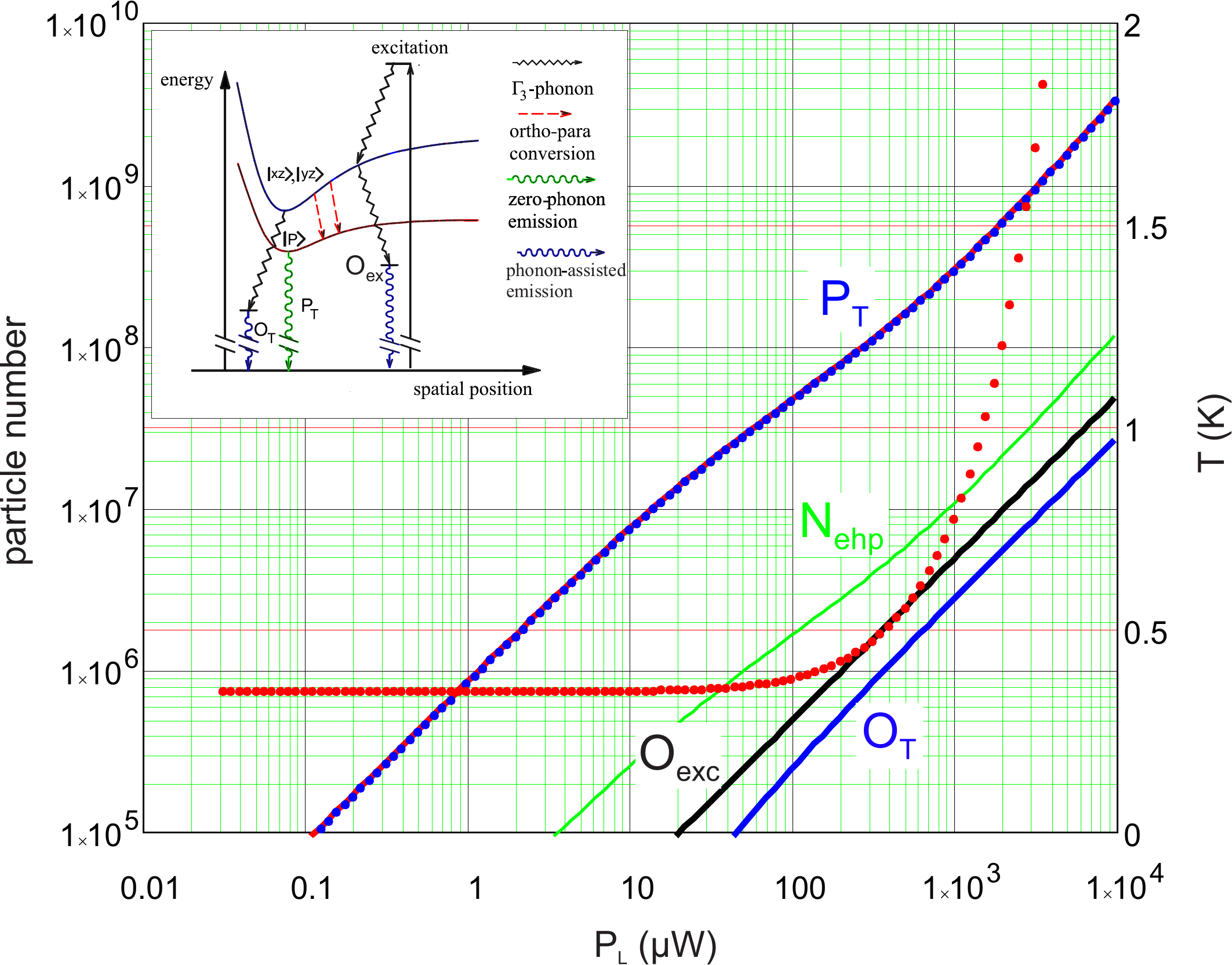}
% \end{minipage}
 \caption{Numbers of para- and orthoexcitons and unbound electron-hole pairs and effective exciton temperature (dotted red line) as a function of excitation laser power as expected from the rate model. The inset schematically shows the laser excitation process to create the orthoexcitons, their conversion into paraexcitons and the luminescence processes to detect the excitons.}
\label{fig:rate_model}
 \end{center}
\end{figure}

We thus obtain the following set of rate equations for the numbers of primarily excited orthoexcitons ($O_{\rm exc}$), trapped ortho- and paraexcitons ($O_{\rm T}$ and $P_{\rm T}$, respectively), and unbound electron-hole pairs $N_{\rm ehp}$, and the effective exciton temperature $T_{\rm eff}$:
\begin{eqnarray}
\frac{dO_{\rm exc}}{dt}&=&N_{\rm 0}\cdot G(t)-\Gamma_{\rm rel}\cdot O_{\rm exc}-\Gamma_{\rm OP}\cdot O_{\rm exc}-\Gamma_{\rm O}\cdot O_{\rm exc} \;,\\
\frac{dO_{\rm T}}{dt}&=&\Gamma_{\rm rel}\cdot O_{\rm exc}-\Gamma_{\rm OP}\cdot O_{\rm T}-2 A_{\rm OO}\cdot O_{\rm T}^2\nonumber \\
&&- A_{\rm OP}\cdot O_{\rm T}\cdot P_{\rm T}-\Gamma_{\rm O}\cdot O_{\rm T}+\frac{1}{2}\Gamma_{\rm rc}N_{\rm ehp}^2\;, \\
\frac{dP_{\rm T}}{dt}&=&\Gamma_{\rm OP}\cdot O_{\rm T} + \Gamma_{\rm OP}\cdot O_{\rm exc}- 2 A_{\rm PP}\cdot P_{\rm T}^2-\Gamma_{\rm P}\cdot P_{\rm T}\nonumber \\ 
&&- A_{\rm OP}\cdot O_{\rm T}\cdot P_{\rm T} +\frac{1}{4}\Gamma_{\rm rc}N_{\rm ehp}^2\;,\\
\frac{dN_{\rm ehp}}{dt}&=& A_{\rm PP}\cdot P_{\rm T}^2 + A_{\rm OO}\cdot O_{\rm T}^2 + A_{\rm OP}\cdot O_{\rm T}\cdot P_{\rm T} - \Gamma_{\rm rc}N_{\rm ehp}^2 -\Gamma_{\rm ehp}N_{\rm ehp}\;,\\\nonumber\\\label{Teff}
\frac{dT_{\rm eff}}{dt}&=& C_{\rm heat}\left[C_{\rm XX} \left(A_{\rm PP}\cdot P_{\rm T}^2 + A_{\rm OO}\cdot O_{\rm T}^2 + A_{\rm OP}\cdot O_{\rm T}\cdot P_{\rm T}   \right)\right. \nonumber\\ 
& &\left. + C_{\rm O} \Gamma_{\rm OP}(O_{\rm T}+O_{\rm exc}) + C_{\rm P}\Gamma_{\rm P}\cdot P_{\rm T}\right] - \Gamma_{\rm cool}(T_{\rm eff}-T_{\rm fin})\, .
\end{eqnarray}
Here, $T_{\rm fin}$ denotes the final temperature to which the exciton gas would relax without additional heating. In an equilibrium situation, this would correspond to the bath temperature. 

To simulate the cw-excitation, we solved the system of rate equations by assuming for $G(t)$ a rectangular shaped excitation pulse of width $\Delta T_{\rm L}=25\, \micro$s and unit pulse area. The number of initially excited excitons $N_{\rm 0}$ is given by 
\be
N_{\rm 0}=\frac{A P_{\rm L}}{\hbar \omega_{\rm L}} \Delta T_{\rm L}\,,
\ee
with the conversion factor from incident laser photons into primarily excited excitons $A=0.45$. All other parameters are taken from reference \cite{schwartz2011}. Since we used the same sample under similar conditions, the calculation allows to obtain the number of para- and orthoexcitons in the trap in dependence on laser power, whereby we guess the accuracy to $\pm 50\%$.  In figure \ref{fig:rate_model} we show a set of typical results. The final temperature $T_{\rm fin}=0.35$~K was chosen to reproduce the experimentally observed temperature dependence in figure \ref{fig:powerspectra}B. The results at high power levels do change somewhat due to the temperature dependence of the Auger process, but not more than $20\%$ by varying $T_{\rm fin}$ from 0.2 K to 0.5 K. 

The number of paraexcitons in the trap at low laser powers turns out to be determined by two parameters, the paraexciton lifetime $1/\Gamma_{\rm P}$ and the fraction of absorbed photons $A$. From figure \ref{fig:rate_model} one can estimate the number of paraexcitons in the trap to be about $8 \cdot 10^6$ for a laser power of $10\, \micro\rm W$. The number of unbound electron-hole pairs would be around $3 \cdot 10^5$. At power levels of $1 \rm mW$, which was the maximum laser power used in the mK experiments, paraexciton numbers of $2\cdot 10^8$ can be realised and the number of electron-hole pairs increases to $10^7$. Assuming thermal equilibrium, this would correspond to densities of about $n_{\rm P}= 2\cdot 10^{16}\,\rm cm^{-3}$ and $n_{\rm ehp}= 10^{14}\,\rm cm^{-3}$.

\subsection{Experimental results}

\subsubsection{Low excitation power}\label{sec:lowp}
\begin{figure}[th]
 \begin{minipage}{16cm}
  \includegraphics[width=0.69\textwidth,angle=-90]{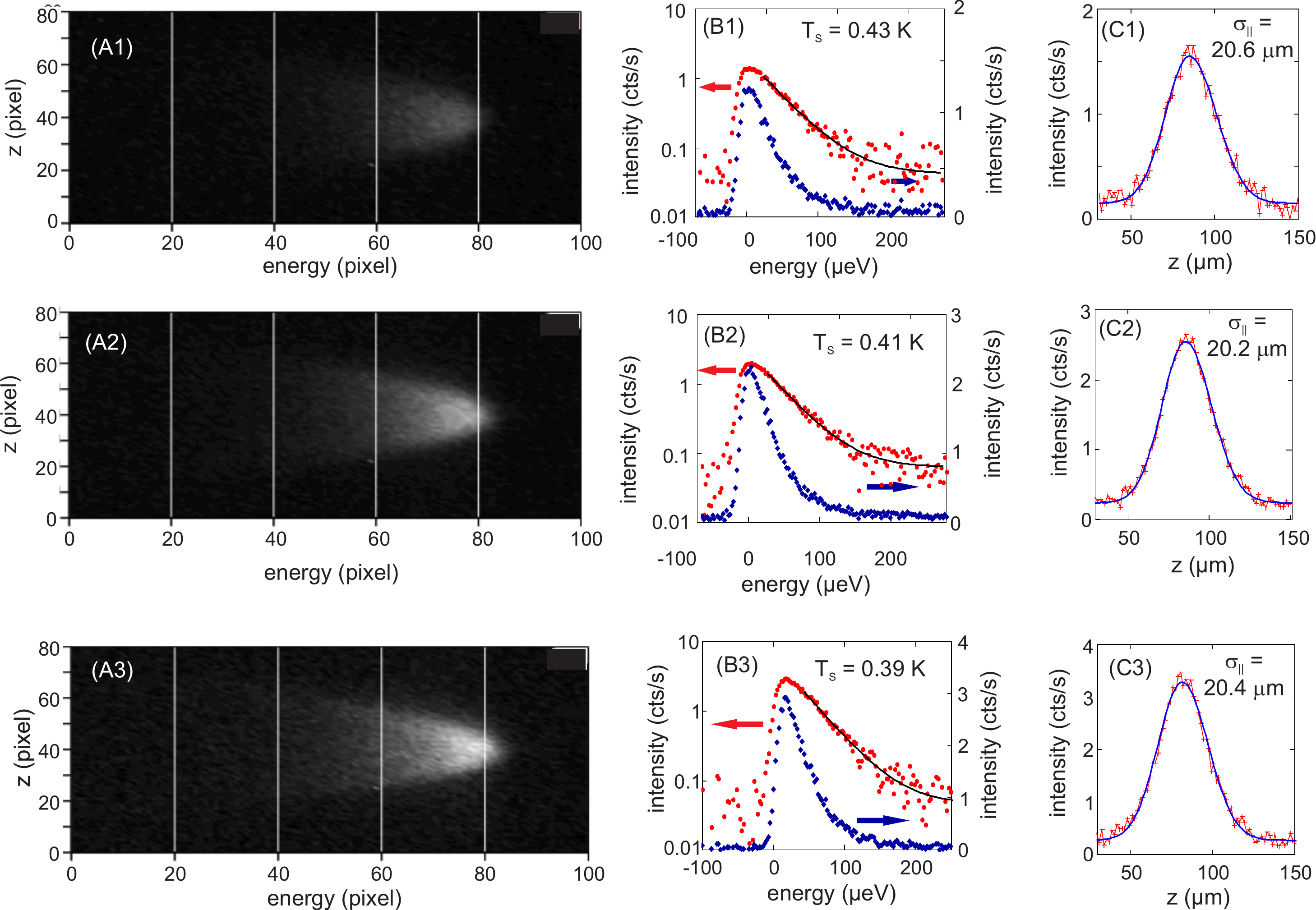}
 \end{minipage}
 \caption{Spectra and spatial luminescence profiles at low excitation power $P_{\rm L}$ with $P_{\rm L}=1.5\,\micro$W (upper row), $P_{\rm L}=\,3\micro$W (middle row), and $P_{\rm L}=\,5\micro$W (lower row). $T_{\rm s}$ denotes the temperature extracted from the high-energy tail of the spectra.}
 \label{fig:spectra_lowp}
\end{figure}

The first experimental results we want to show were obtained at rather low laser excitation powers in the range between $1$ and  $5\, \micro\rm W$ with high spectral resolution (figure \ref{fig:spectra_lowp}).  Excitation was performed via the phonon-assisted absorption of the orthoexcitons with energy slightly below the bulk exciton gap as described in section \ref{sec:expsetup}. Orthoexcitons were converted rapidly into paraexcitons and relax down to the bottom of the trap.   

\begin{figure}[b]
\begin{center}
 \begin{minipage}{14cm}
  \includegraphics[width=\textwidth]{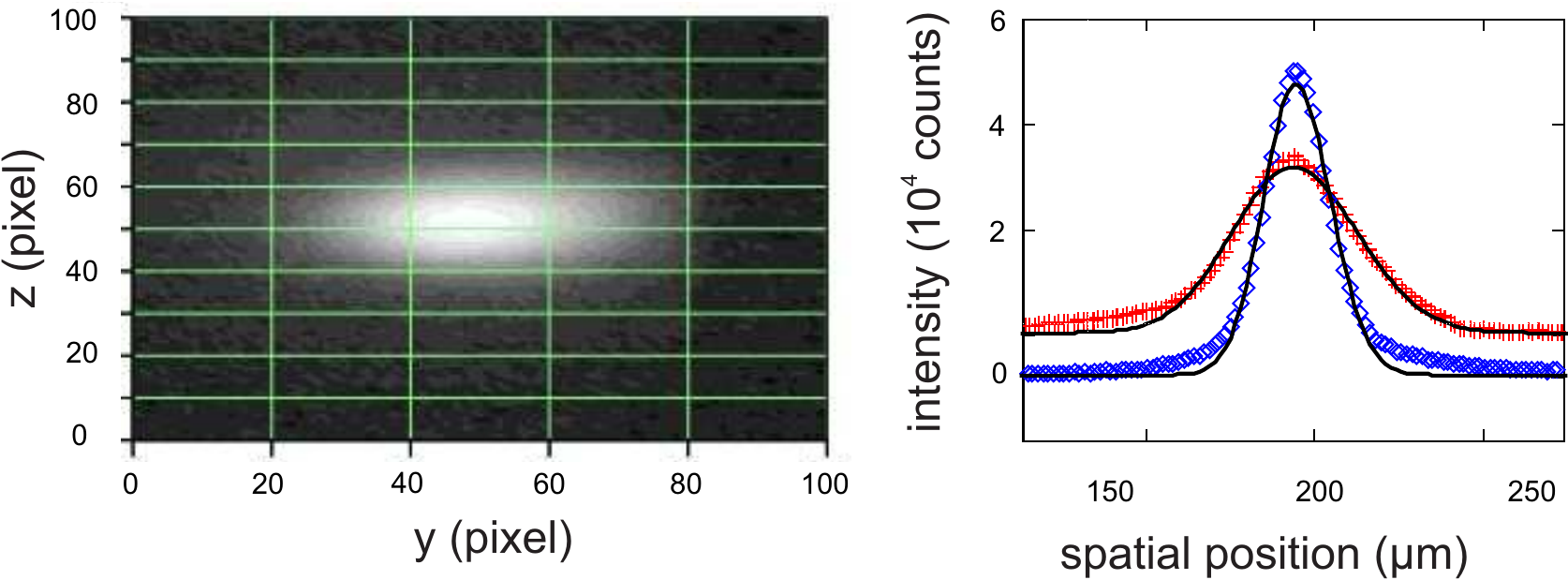}
 \end{minipage}
 \caption{left panel: Spatial image of the exciton cloud at low excitation power of $P_{\rm L}=30\,\micro$W. Right panel: Marginal intensity distributions along $z$ (blue circles) and $y$ (red circles) with fits by gaussians according to equation \ref{eqn_distr}.}
 \label{fig:spatial_im}
\end{center}
\end{figure}

From the $z$-resolved spectra (panels A) we obtained the $z$-integrated spectra (panels B) and the $z$-profiles (panels C). As already could be seen in the $z$-resolved spectra, there is no change in both the spectral and spatial distribution, only an increase in the overall intensity which is almost proportional to the laser power. This demonstrates that we are in the linear excitation regime and that bimolecular decay processes are not important. The high-energy tail of the spectra can be fitted quite well by a Boltzmann distribution with an effective exciton temperature of about $0.4$ K. A two-dimensional spatial image of the exciton cloud was obtained by using the spectrometer in the subtractive mode, the output stage at zero wavelength and by setting the intermediate slit of the subtractive stage to just let pass the emission from the paraexcitons in the trap. A typical example is shown in figure \ref{fig:spatial_im}.
The spatial profiles are described by simple Gaussian distributions 
\be  \label{eqn_distr}
I(y,z)= n_{\rm 0} \exp\left(-\nicefrac{y^2}{\sigma^2 _{\perp}}\right)\exp\left(-\nicefrac{z^2}{\sigma^2 _{\|}}\right)
\ee
with width parameters $\sigma_\perp$ and $\sigma_\parallel$ of about $20\, \micro\rm m$.
%ure\ref{fig:spatial_im}.

According to the theoretical predictions in thermal equilibrium (see section \ref{theory-1}), the width parameters are related to the curvature of the trap $\alpha$ and temperature $T$ by  
\be \label{size-para}
\sigma _{\perp,\|}= \sqrt{\frac{k_{\rm B} T}{\alpha_{\perp,\|}}}\, .
\ee
However, the ratio of the resulting width parameters does not fit the ratio of the potential curvatures as obtained from the strain calculation. It comes out about 30\% larger. As origin of this broadening we identified a small vibrational motion of the sample in $y$-direction of the order of $10\, \micro\rm m$ which could be directly observed by imaging a submicron sized needle. For a quantitative theoretical model of the light emission of the excitons, this effect has to be taken into account.
   
Using the curvature parallel to the $z$-direction from table 1, and subtracting the spatial blurring due to diffraction of our optical imaging setup ($7\, \micro\rm m$), we obtain from the spatial profiles temperatures of about $0.45$ K, in good agreement with those from the spectra. On a first sight, we would conclude that the excitons are in a state of quasi-thermal equilibrium, albeit with a temperature which is more than an order of magnitude larger than the bath temperature, which was in these experiments around 38 mK. 
There are two possible explanations of this difference:
\begin{enumerate}
\item
The heating of the exciton cloud due to the excess kinetic energy of about 8 meV in the ortho-para conversion process or due to the exciton-exciton annihilation process. The excess electronic energy is in both cases converted into phonons and finally increases the temperature of the lattice, which is unknown in the present experiment, however.  This heat has to be transferred into the surrounding He bath, where it would lead to an increase in the temperature of the mixing chamber, which could be measured during the experiment. 

Indeed we observe this heating effect at higher laser powers. As shown in figure \ref{fig:powerspectra}, the effective exciton temperature and the temperature of the mixing chamber rise from their initial values very steeply above $P_{\rm L} = 1000\,\micro\rm W$. Presuming a well-defined relation between the lattice temperature of the crystal and that of the mixing chamber, we have to conclude from these data that, at low laser powers, the heating effect is not present and the lattice temperature should be close to that of the He-bath at very low laser powers. This conclusion is in agreement with the results of the rate model presented in the foregoing section, too. There the heating of the exciton cloud is starting at laser powers well above $100\, \micro\rm W$ and is not present at low powers.

\begin{figure}[hb]
 \begin{minipage}{16cm}
  \includegraphics[width=0.41\textwidth,angle=-90]{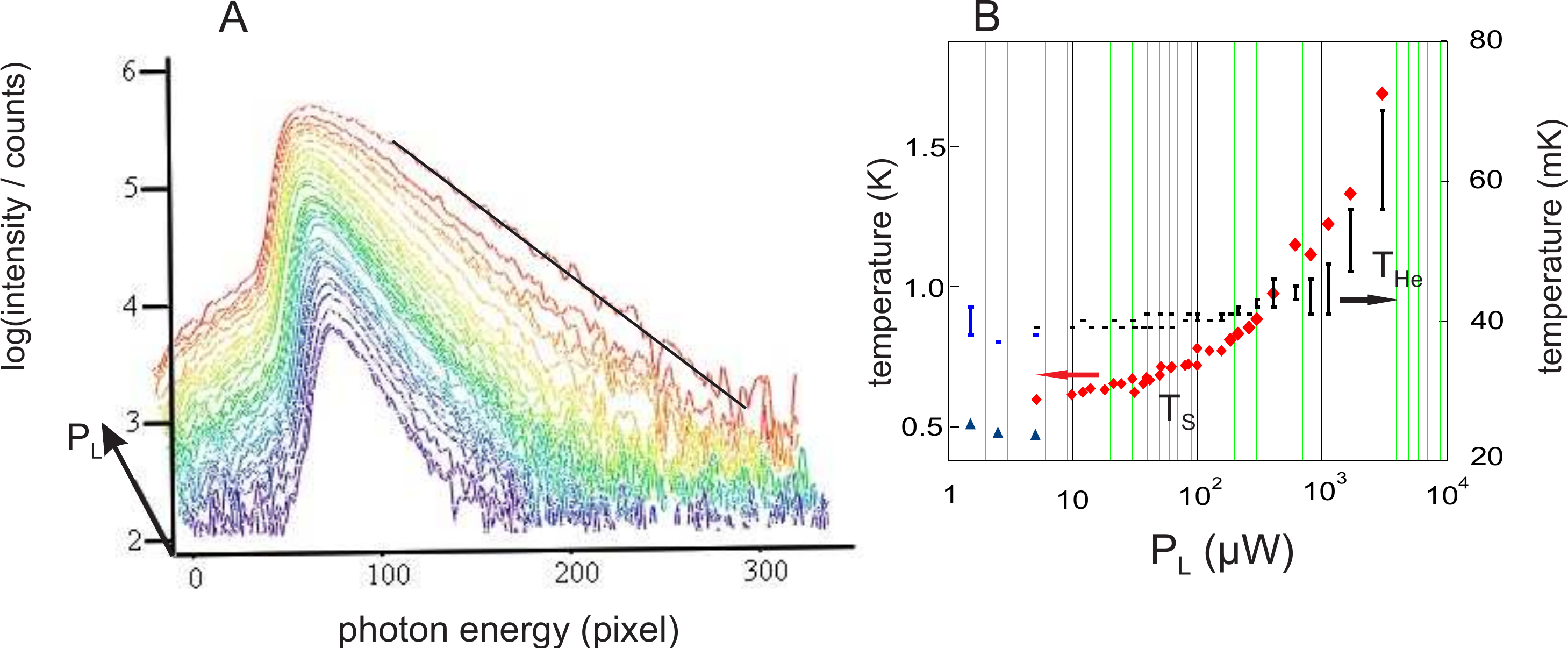}
 \end{minipage}
 \caption{Panel A: Series of spectra taken at increasing laser power $P_{\rm L}$ from 5 $\micro$W to 3 mW in a logarithmic intensity scale. The spectral position is given in CCD-pixels, 1 pixel corresponds to $2.5\,\micro$eV.
 Panel B: Spectral temperature $T_{\rm s}$ as obtained from a Boltzmann fit of the high-energy tail of the spectra shown in panel A (left ordinate) and the temperature of the mixing chamber $T_{\rm He}$ (right ordinate).}
 \label{fig:powerspectra}
\end{figure}

\item
The discrepancy of the lattice temperature $T_{\rm latt}$ and that of the exciton cloud may result because the excitons do not reach global thermodynamic equilibrium during their finite lifetime. This standard point of view means that, due to elastic exciton-exciton scattering, a quasi-thermal equilibrium state is established very fast with a temperature above that of the lattice. Hereafter, the exciton gas cools down to the lattice temperature by emission of acoustic phonons. The time scale of this process increases with lower lattice temperature. In an external potential provided by the trap, the excitons are additionally driven by force and drift terms into the potential minimum, as described by the Boltzmann equation \cite{ashcroft}. The detailed balance of these processes is determined by the interplay between the various exciton relaxation mechanisms. 

In a recent paper, we published a detailed numerical simulation by solving the Boltzmann transport equation for excitons in a potential trap under pulsed optical excitation  considering phonon relaxation and exciton-exciton annihilation \cite{som2012}. The simulation indeed showed a strong non-equilibrium situation. While the local exciton energy distribution reached, within the exciton lifetime of several 100 ns, an equilibrium with the lattice down to temperatures above 300 mK, the spatial distribution, which is governed by the force and drift terms, remained much broader than in thermal equilibrium. For the case of still lower lattice temperature, which was not considered in \cite{som2012}, we obtain locally thermodynamic quasi-equilibrium also at temperatures down to 50 mK by including elastic exciton-exciton scattering as is shown in detail in \ref{app:relax}.  Albeit in this range the exciton temperature does not come down to the lattice temperature within the exciton life time, it is much lower than the observed values of $T_{\rm s}$. According to these results, in our experiments we should expect a non-equilibrium situation in which the excitons locally are in quasi-thermal equilibrium with a temperature $T_{\rm X}$, which we assume to be the same at all points in the trap. Globally, the exciton distribution will be quite different, but still can be described approximately by a Gaussian dependence on position.

\end{enumerate}

Considering the effect of such a situation on the experimentally measured luminescence spectra, we recall that only excitons near $k_{\rm 0}$ participate in the light emission, where $\mathbf{k}_{\rm 0}$ is the wave vector of the intersection of the photon and exciton dispersions. Its modulus is given by $k_{\rm 0}= E_{\rm gX} n/\hbar c$, with $E_{\rm gX}$ being the excitonic band gap, $n$ the refraction index, and $c$ the vacuum velocity of light. This means, we observe the exciton distribution function 
$f_{\rm eq}(\mathbf{k},\mathbf{r})$ 
only at one point in local $k$-space. 
Furthermore, due to energy conservation, at a certain spectral position $\hbar\omega$ we only observe those excitons, which are at that spatial position in the trap, where the trap potential $V_{\rm ext}(\mathbf{r})=  \hbar\omega -E_{\rm 0}-\varepsilon_{\rm 0}$. Here, $E_{\rm 0}$ is the energy of the trap bottom and $\varepsilon_{\rm 0}$ is the kinetic energy of the excitons at $k_{\rm 0}$. If we have a spatial distribution
\begin{equation}
n(r) \propto \exp(-r^2/\sigma^2)\,,
\end{equation}
this yields a spectral distribution of the form
\begin{equation}
I(\hbar\omega) \propto \exp(-(\hbar\omega -E_{\rm 0}-\varepsilon_{\rm 0})/(\alpha \sigma^2))\,,
\end{equation}
which suggests a Boltzmann distribution with an effective temperature $T_{\rm s}=\alpha \sigma^2/k_{\rm B}$. Then the spectral distribution of luminescence intensity in reality describes the spatial distribution and vice versa. Therefore, it has to result in the same effective exciton temperature, which is also a consistency check for the trap potential. Consequently, we will designate this temperature as the {\it spatial temperature} $T_{\rm s}$ of the exciton cloud. Obviously, $T_{\rm s}$ corresponds to $T_{\rm eff}$ of the rate model, cf.\ equation (\ref{Teff}).
On the other hand, it is not possible to extract the local effective exciton temperature $T_{\rm X}$ from the intensity of the zero-phonon transition. This would require a spectroscopic technique which is sensitive to the local exciton distribution like the transitions between 1S and 2P exciton states (see \cite{klingshirn2005,gonokami2005}). Hence, at the present stage of measurements, all we can state is that $T_{\rm X}$ may differ considerably from both the lattice temperature, which will set a minimum value, and the spatial temperature, which will set a maximum value, as in the latter case we would have global equilibrium.

\subsubsection{Power dependent experiments}
\begin{figure}[b]
 \begin{center}
  \includegraphics[width=0.9\textwidth]{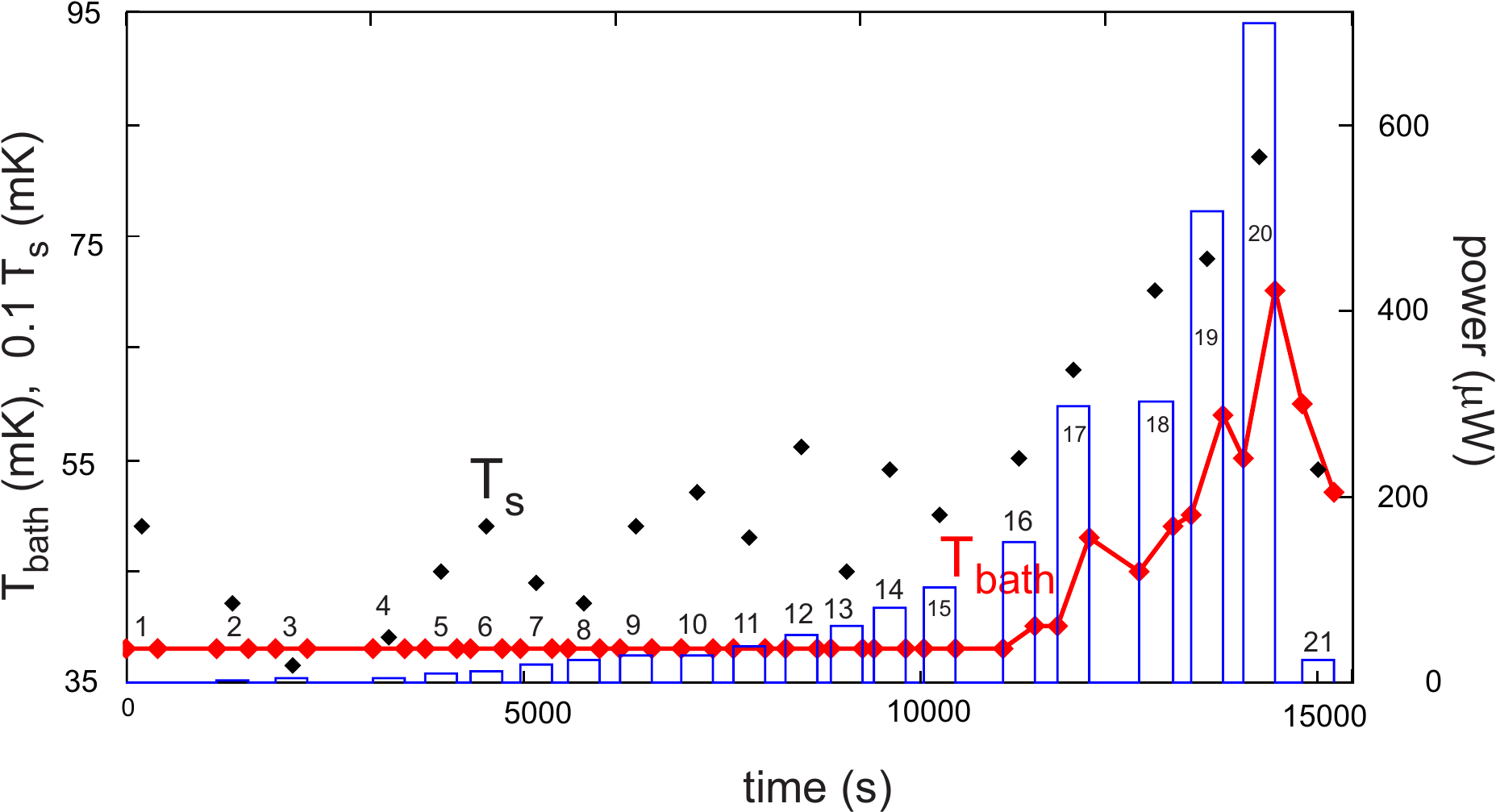}
 \end{center}
 \caption{Time profiles of laser power (blue bars) and temperature of the mixing chamber (red diamonds), and spatial temperature (black diamonds) for the measurements shown in figure \ref{fig:series_im}. The red line is a guide for the eye.}
 \label{fig:power_temp}
\end{figure}

In the following we will present a series of measurements under controlled conditions of excitation and cooling times such that we are able to influence the thermodynamic state of the system in order to optimise any possible phase transition. The laser power ranged from below $1\,\micro\rm W$ to about $700\, \micro\rm W$. For each measurement we noted the temperature of the mixing chamber  (He-bath temperature $T_{\rm bath}$) at the beginning and at the end. Between the measurements, the laser beam was blocked, so that the crystal could cool down (see figure \ref{fig:power_temp}).

In order to obtain the best signal-to-noise ratio, the ICCD camera was operated in the photon counting mode. Here, the CCD-chip is read out after a short exposure time ($T_1$) and the positions of the single photon peaks are determined and stored. A single photon is identified, if the signal falls between a lower and an upper discriminator level, which were adjusted so that only single photon signals fall in this range. For a number $N_{\rm acc}$ of accumulations, the data are summed up in the memory. Advantages of this method are that signals due to cosmic rays are easily detected and eliminated and that one can estimate the error of the intensity signal due to the Poissonian statistics as the square root of the number of photons collected at each pixel. Due to the Poissonian statistics, the probability of detecting $n$ photons is given by
\be
P_n= \frac{\lambda^n}{n!}\exp(-\lambda)\,,
\ee
where $\lambda$ is the average photon number for each exposure. As we detect only single photons and throw away all higher photon number events, $\lambda$ should be much smaller than one. However, for moderate values, we can obtain the true number of detected photons, $\lambda N_{\rm acc}$ from the measured number $N_{\rm det}$ by applying the correction formula
\be
\lambda N_{\rm acc} = -\ln\left(1-\frac{N_{\rm det}}{N_{\rm acc}}\right)\,.
\ee
 
\begin{figure}[h]
\begin{center}
 \begin{minipage}{16cm}
  \includegraphics[width=0.56\textheight,angle=-180]{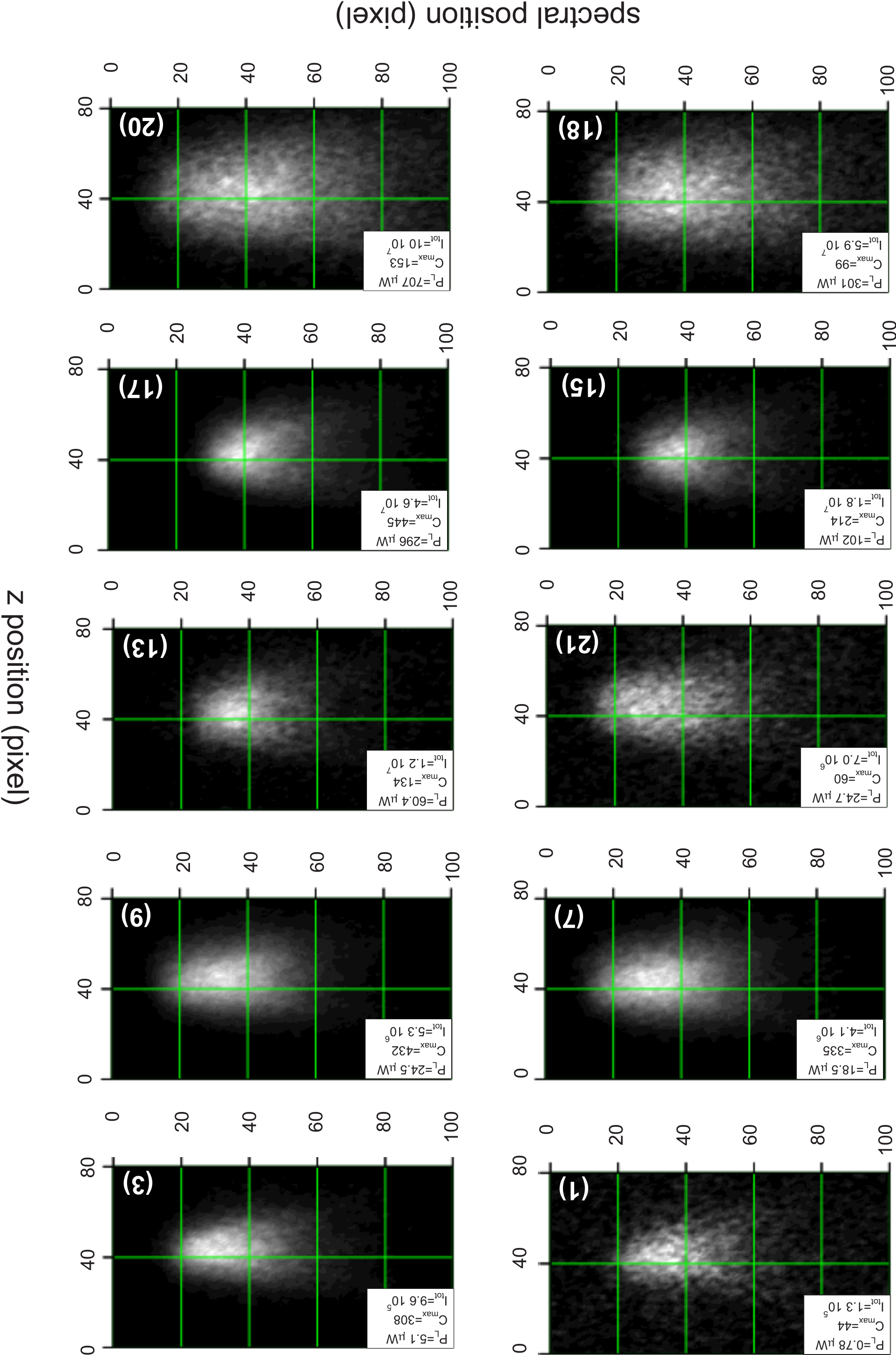}
 \end{minipage}
 \caption{Series of $z$-resolved luminescence spectra for different excitation powers $P_{\rm L}$ and bath temperatures (see figure \ref{fig:power_temp}). $C_{\rm max}$ denotes the peak value and $I_{\rm tot}$ the totally integrated intensity. The numbers in parentheses denote the measurement numbers, cf.\ figure \ref{fig:power_temp}.}
 \label{fig:series_im}
\end{center}
\end{figure}

For the measurements of the $z$-resolved luminescence spectra shown in figure \ref{fig:series_im} we have $T_1= 0.5$ seconds and $N_{\rm acc}=400$. In order to keep the maximum number of detected photons well below $N_{\rm acc}$, we inserted neutral density filters before the entrance slit of the monochromator. Their transmission factors have been determined by direct intensity calibration as $D1=0.314$ and $D2=0.0333$. 

In figure \ref{fig:series_im}, each measurement is numbered according to its position in time given in figure \ref{fig:power_temp}. For each image, we specify the laser power, the maximum number of counted photons, which allows to judge the signal-to-noise ratio, and the totally integrated intensity. 
At first sight, all the images look very much the same, despite the three orders of magnitude increase in laser power. In particular, we observe neither the sharp spatial and spectral peak at the bottom of the trap, which is expected as a signature for BEC in case of an ideal Bose gas, nor the predicted flattening of the emission along the $z$-direction in case of an interacting Bose gas \cite{stolz2010}. In this respect, the measurements shown are similar to those reported in all earlier work \cite{wolfe1986,gonokami2011}. A closer look, however, reveals subtle changes of the shape of the spectra depending on the laser power. For power levels below $30\, \micro \rm W$, the $z$-resolved spectrum of the exciton cloud shows no significant changes -- here note especially measurements 9 and 21, which were taken under very different He-bath temperatures (see figure \ref{fig:power_temp}). For power levels above $30\, \micro \rm W$, the images change gradually by showing a spectral narrowing and a slight blue shift. These changes suddenly come to a stop between measurements 17 and 18, with the images from now on both spatially and spectrally much broader and resembling those taken at low powers.  The latter data have been taken at almost the same laser power of $300\, \micro\rm W$, but the He-bath temperature was significantly different. Measurement 17 started at $T_{\rm bath}=42\,\rm mK$. During the measurement, the laser power resulted in a heating up to $T_{\rm bath}=49\,\rm mK$. In contrast, measurement 18 already started at $T_{\rm bath}=45\,\rm mK$  and stopped at $T_{\rm bath}=50\,\rm mK$. This temperature variation of the mixing chamber should directly reflect the actual lattice temperature $T_{\rm latt}$ of the Cu$_2$O crystal. At the beginning of measurement 17, $T_{\rm latt}$ must have been quite close to that of the bath while at the start of measurement 18, $T_{\rm latt}$ has only cooled down to midway between start and end of measurement 17. This difference in lattice temperature is reflected in a concomitant difference in the spectral temperature which is $T_{\rm s}=0.62$ K for measurement 17 and $T_{\rm s}=0.72$ K for measurement 18. The totally integrated intensities of both measurements are expected to be similar due to the same laser power, however, that of measurement 18 is about 30\% larger. 

We can resolve this puzzle by assuming that the difference between both measurements comes from the existence of a Bose-Einstein condensate of excitons in measurement 17, while measurement 18 describes a normal exciton cloud at temperatures above the critical temperatures for BEC for the actual number of excitons in the trap, which is the same in both measurements. Then the less intensity in 17 would be the result of the decreased number of thermalised excitons, since according to \cite{stolz2010} and anticipating the discussion in section \ref{theory-1}, excitons in the condensate are in the ground state ($k=0$), and, therefore, could not emit light with wave vector $k_{\rm 0}$ at all (as in a homogeneous system), or strongly suppressed (as would be the case in the trap). 
 
\begin{figure}[h]
 \begin{minipage}{7.8cm}
  \includegraphics[width=\textwidth]{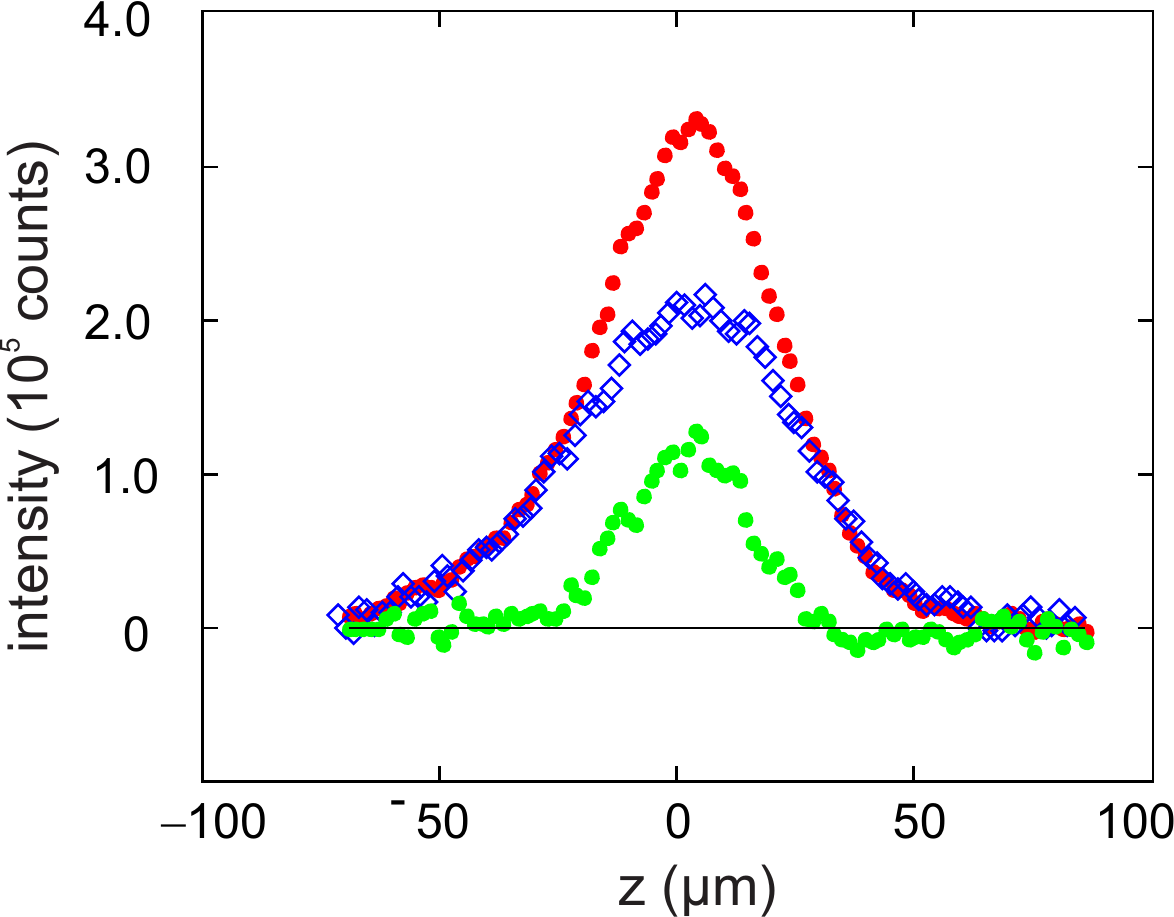}
 \end{minipage}
 \begin{minipage}{7.8cm}
  \includegraphics[width=\textwidth]{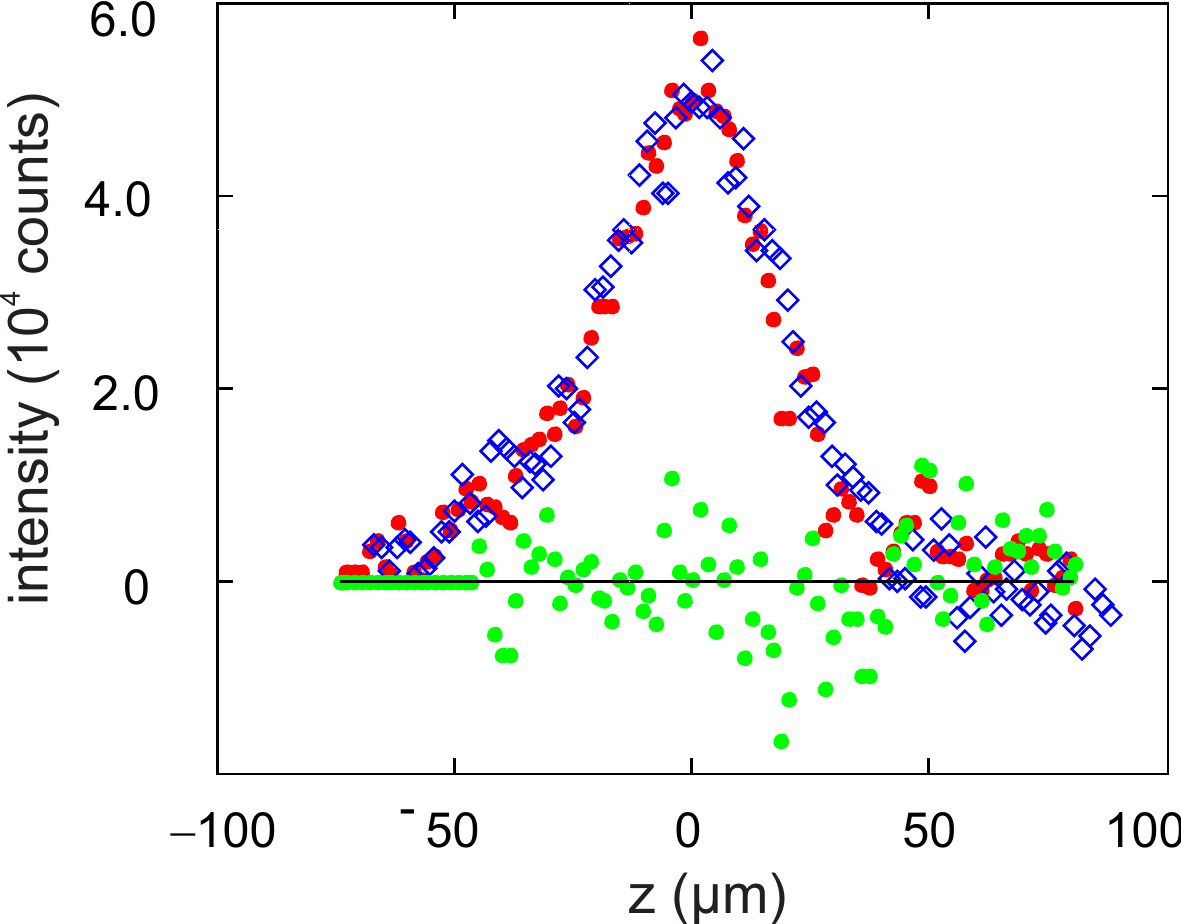}
 \end{minipage}

 \caption{Comparison of $z$-profiles of measurement 17 and 18 (left panel) and 9 and 21 (right panel). Each pair was taken at the same laser power, but different bath temperature. For details see text.}
 \label{fig:comp_820_1617}
\end{figure}

\begin{figure}[h]
\begin{center} 
 \begin{minipage}{11cm}
  \includegraphics[width=\textwidth]{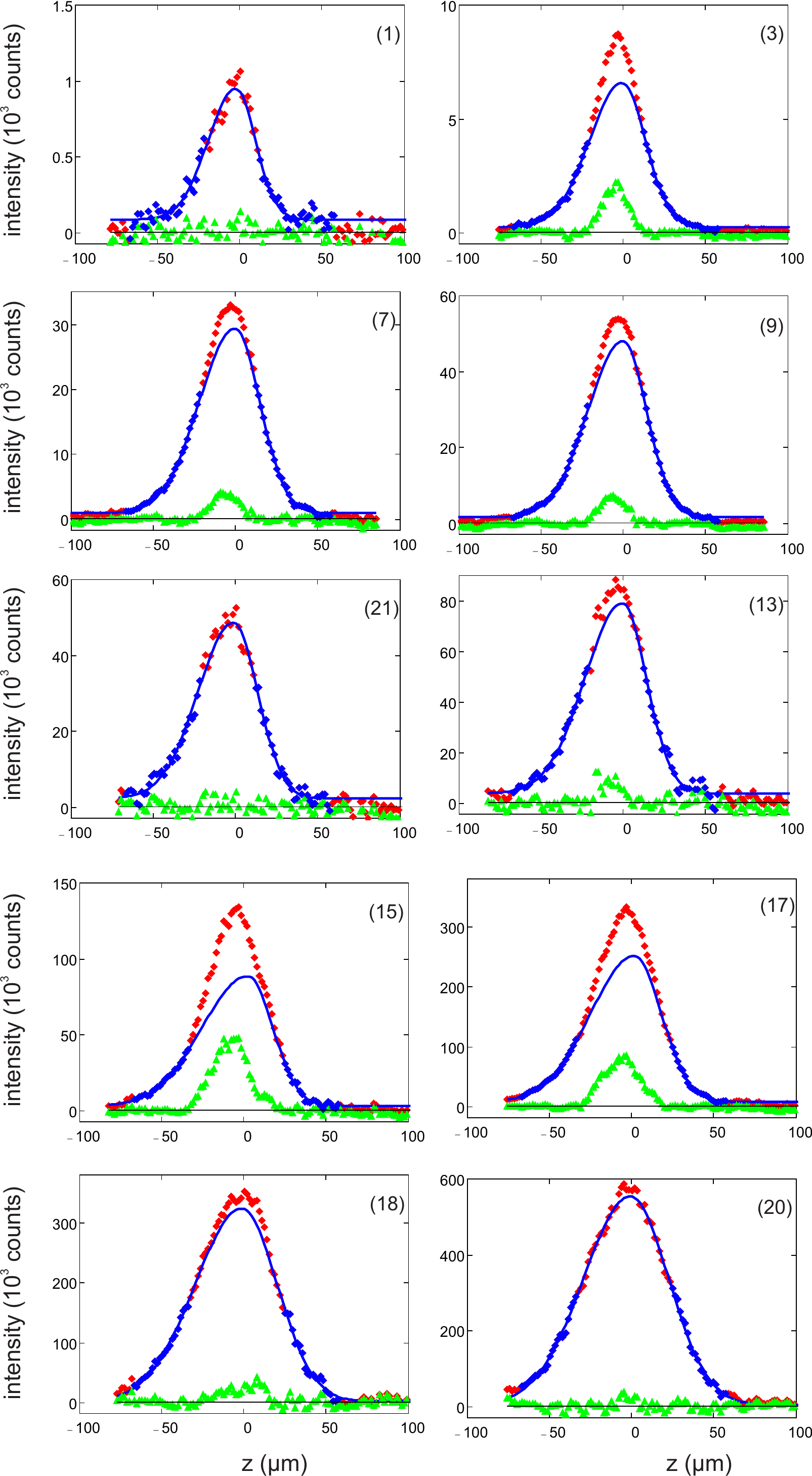}
 \end{minipage}
 \caption{Comparison of $z$-profiles with fits by a Gaussian. Red and blue diamonds: total measured profiles, solid blue lines: Gaussian fit of the total profiles, blue diamonds: data points used for the fit, green triangles: difference between total profile and fit. The numbers in parentheses denote the measurement numbers, cf.\ figure \ref{fig:power_temp}.}
 \label{fig:comp_z}
\end{center}
\end{figure}

Thus, we propose the following scenario for measurement 17: In the beginning, we have a system with a large fraction of the excitons in a condensate, with a reduced luminescence intensity. As time increases, the crystal heats up and the condensate fraction is reduced. When the lattice temperature reaches the critical temperature, we have a situation just like in measurement 18. Therefore, after being integrated over the exposure time $T_1$, the $z$-resolved spectra should look the same. In a first approximation, the image of 17 is a superposition of that of measurement 18 reduced by a factor $\gamma$ and that of a system with a condensate. To demonstrate that this is indeed the case, we compare in figure \ref{fig:comp_820_1617} in the left panel the $z$-profiles for both measurements. By assuming $\gamma=0.6$, the wings of both profiles do coincide exactly, which is clearly visible in the difference profile. The difference around the trap centre, however, has a characteristic, non-Gaussian line shape resembling that of a condensate (compare figure \ref{fig:comp_z}). A similar comparison of measurements 9 and 21, both taken at a laser power of $25\,\micro\rm W$ (right panel of figure \ref{fig:comp_820_1617}) shows no difference in the $z$-profiles within the statistical errors. 

While such a comparison of different measurements seems to be intuitively correct, there are several objections against this procedure. First, the measurements we compare have quite different values for the spatial temperature and thus different spatial extension in the trap. Second, we cannot exclude a priori that, e.g., measurement 18 has a condensate as well.
To overcome these difficulties, we have to look for a way to analyse each measurement for itself. Theoretical studies of the thermodynamics of an interacting Bose gas of excitons \cite{stolz2010} have shown that even in case of a BEC both the density of excitons and the spectrally integrated luminescence intensity outside the condensate region closely follow a Gaussian distribution reflecting the temperature of the exciton gas. Therefore, by fitting a Gaussian only to the wings at larger $z$ of the intensity profile, we should have access to the distribution of thermal excitons only. If there is no condensate present, this Gaussian should also describe the intensity profile in the centre of the trap. On the other hand, any deviation of the measured profile from this curve is a clear indication of an additional luminescing component in the trap. If the line shape of this component is non-Gaussian, this would clearly indicate the existence of a condensate. 

\begin{figure}[h]
\begin{center}
 \begin{minipage}{10cm}
  \includegraphics[width=\textwidth]{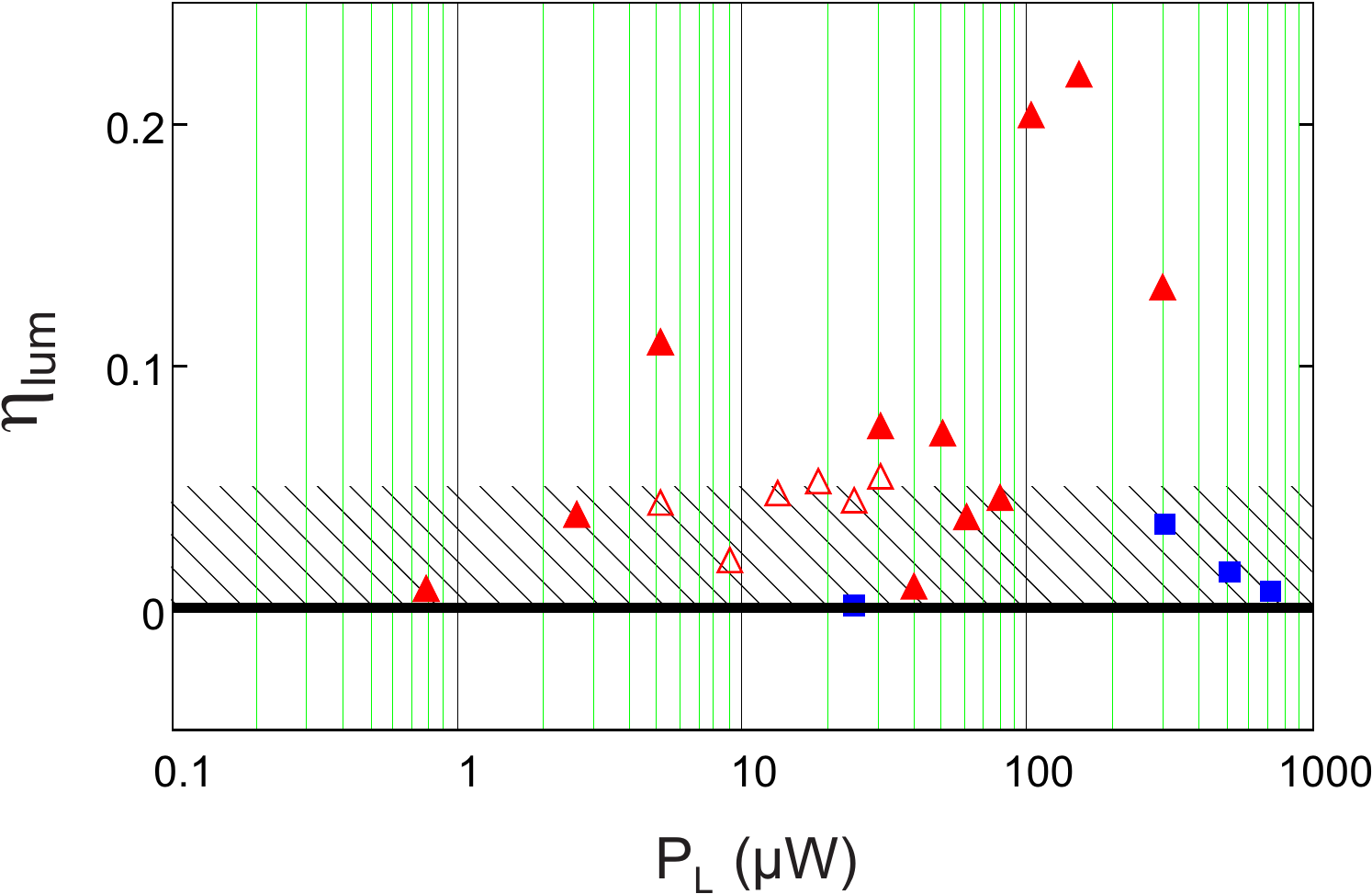}
 \end{minipage}
 \caption{Power dependence of the ratio between integrated intensities of the condensate and of the total band. Filled red triangles: measurements 1-3 and 10-17, open red triangles: measurements 4-9, blue squares: measurements 18-21. Between measurements 3 and 4, and between measurements 9 and 10, there was a change of the neutral density filter, cf. figure \ref{fig:power_temp}. The shaded band denotes an area where the contribution is too small for a significant proof of a condensate.}
 \label{fig:condlumfrac}
\end{center}
\end{figure}

For the measurements shown in figure \ref{fig:series_im}, the results of such a fitting procedure are shown in figure \ref{fig:comp_z}. The red and blue diamonds are the experimental points, whereby the blue diamonds mark those intensity points that have been used in the fitting. The blue line gives the shape of the Gaussian. In order to reproduce the obviously different line shapes for positive and negative $z$ values due to the Morse-type potential of the trap in $z$-direction, we used for the fit an asymmetric Gaussian function of the form
\be
S(z)\propto \mathrm{exp}\left(-\frac{(z-z_{\rm 0})^2}{(\Theta(z_{\rm 0}-z)\sigma_- +\Theta(z-z_{\rm 0})\sigma_+)^{2}}\right)\,.
\ee
For all measurements, the ratio of $\sigma_+/\sigma_-$ is obtained as 1.5, in consistence with the shape of the potential.
The number of data points for the fitting was chosen for each set of data such that the average error per point becomes minimum. If we chose less points, the statistical error due to noise will increase, if we chose more points, the systematic error increases if the profiles contain an additional contribution. The green triangles show the difference of the experimental points and the Gaussian fits.
In total, the results of the procedure substantiate our previous analysis. The measurements at very low power (1) and at high bath temperature (20,21), where we expect that there is no condensate present, indeed can be fitted completely by a single Gaussian. Measurements 15 and 17 show clearly a bimodal distribution and thus a strong condensate component. However, in measurement 18 a small additional contribution remains. To show systematically the results of the fit, we have plotted in figure \ref{fig:condlumfrac} the ratio of the green component to that of the overall intensity. Looking more closely on the results shown in figure \ref{fig:condlumfrac}, we can identify at least four sets of data with a strong condensate contribution, while most of the other measurements show a contribution below 0.05, which we will consider not as significant. The occurrence of a condensate in the power range between 60 and 300 $\micro \rm W$ is understandable in view of the rather high number of paraexcitons in the trap which, according to the rate model (figure \ref{fig:rate_model}) is about $4 \cdot 10^7$ at $100\, \micro \rm W$. The appearance of a condensate at the low power of $5\, \micro \rm W$ is rather surprising and requires further considerations. Indeed, the theory presented in the next section will show that, under the conditions of our experiments (bath temperature 38 mK and no heating of the sample), condensation will take place at exciton numbers as low as $4 \cdot 10^6$ (see figure \ref{fig:nctstx}).   

\begin{figure}[h]
\begin{center}
 \begin{minipage}{10cm}
  \includegraphics[width=\textwidth]{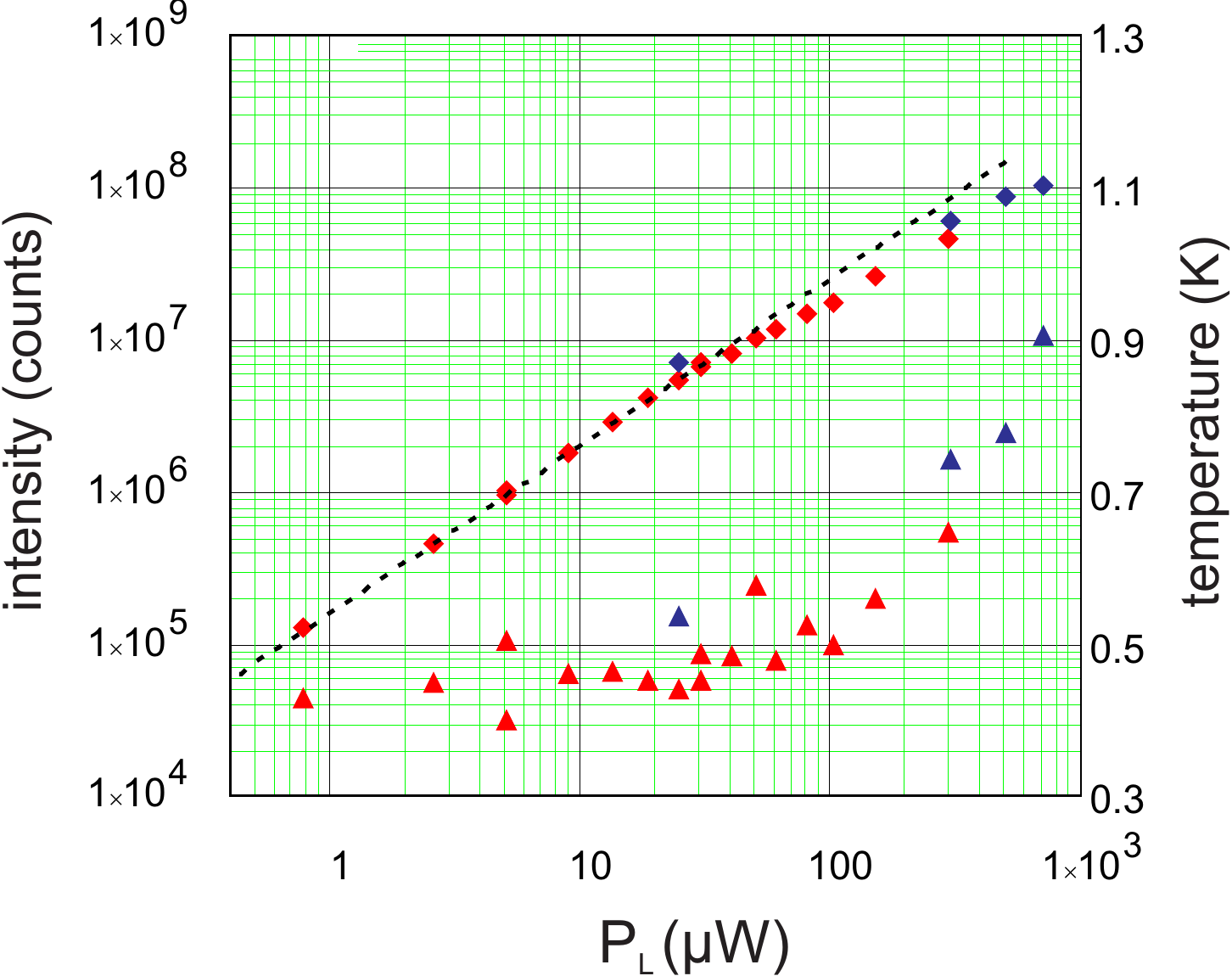}
 \end{minipage}
 \caption{Dependence of the totally integrated intensity (squares) and spatial temperature (triangles) on the laser power. The points with blue colour mark measurements 18-21. The dashed line gives a power dependence according to $P_{\rm L}^{b}$ with $b=1.1$ which describes the low power regime quite accurately. }
 \label{fig:isotherm2510}
\end{center}
\end{figure}

\begin{figure}[h]
\begin{center}
% \begin{minipage}{15cm}
  \includegraphics[width=0.8\textwidth]{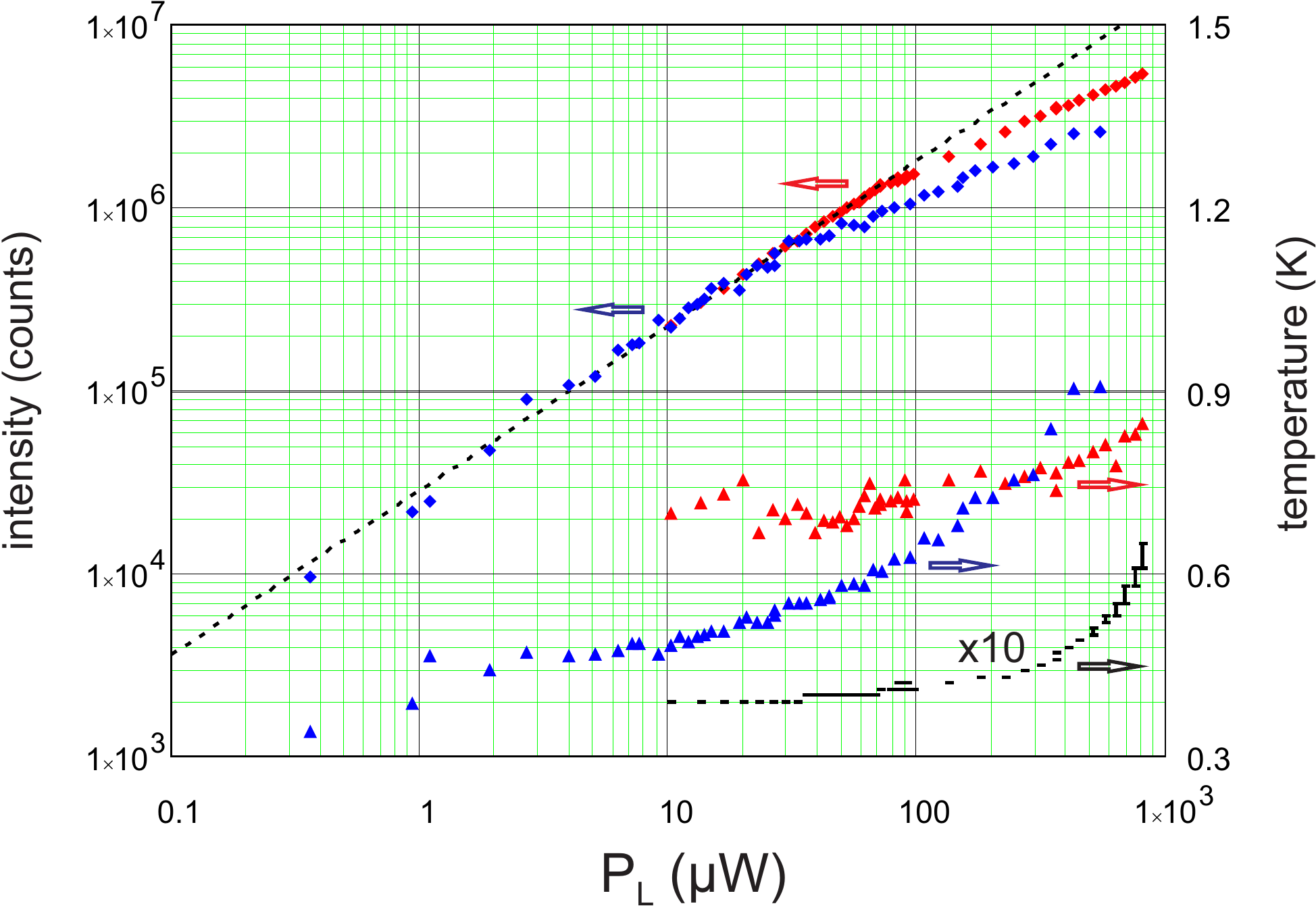}
% \end{minipage}
 \caption{Dependence of the totally integrated intensity (squares), spatial temperature (triangles), and start and final bath temperature during the measurement (black bars) on the laser power for two different sets of data.}
 \label{fig:isothermen_all1}
\end{center}
\end{figure}

Here we stress that the luminescence fraction $\eta_{\rm lum}$ shown in figure \ref{fig:condlumfrac} is different from the fraction of particles in the condensate $\eta_c=N_c/N_{\rm tot}$. If $f_{\rm lum}$ denotes the ratio of the luminescence efficiency of excitons in the condensate to that of excitons in the thermal cloud (see section 3.2), the relation is given by
\be \label{eq:condlumfrac}
\eta_{\rm lum}=\frac{f_{\rm lum}\eta_c}{1-\eta_c+f_{\rm lum}\eta_c}\,.
\ee
At present, $f_{\rm lum}$ is neither known experimentally nor theoretically (see section 3.2), however, it must be smaller than one because otherwise we would not observe a kink in the luminescence dependence on laser power (see figures \ref{fig:isotherm2510} and \ref{fig:isothermen_all1}) so that $\eta_c > \eta_{\rm lum}$.

Our interpretation is further substantiated by the dependence of the totally integrated intensity on the excitation laser power which is shown in figure \ref{fig:isotherm2510}. 

At low power, the dependence is almost linear, indicating that decay processes by exciton-exciton collisions are not important. Actually, the dependence is even slightly superlinear which might be due to the effect of increasing homogeneous broadening at higher exciton numbers. At power levels of about $60\,\micro \rm W$, we clearly observe a kink resulting in a weaker slope at higher power levels. This is just the behaviour predicted by the theory in section \ref{sec:inegratedlum}. The critical power at the kink is in full agreement with the conclusion drawn from figure \ref{fig:condlumfrac}, where we observed the onset of condensation at powers above $60\,\micro \rm W$.  This dependence is abruptly changed for the measurements 18-21 in agreement with the vanishing of the condensate. At the high exciton numbers (under these conditions, we expect $N\simeq 1\cdot 10^8$ according to figure \ref{fig:rate_model}) we have a substantial effect of the two-body decay resulting in a sublinear behaviour, see the data marked by blue symbols.

\begin{figure}[h]
\begin{center}
% \begin{minipage}{15cm}
  \includegraphics[width=0.7\textwidth]{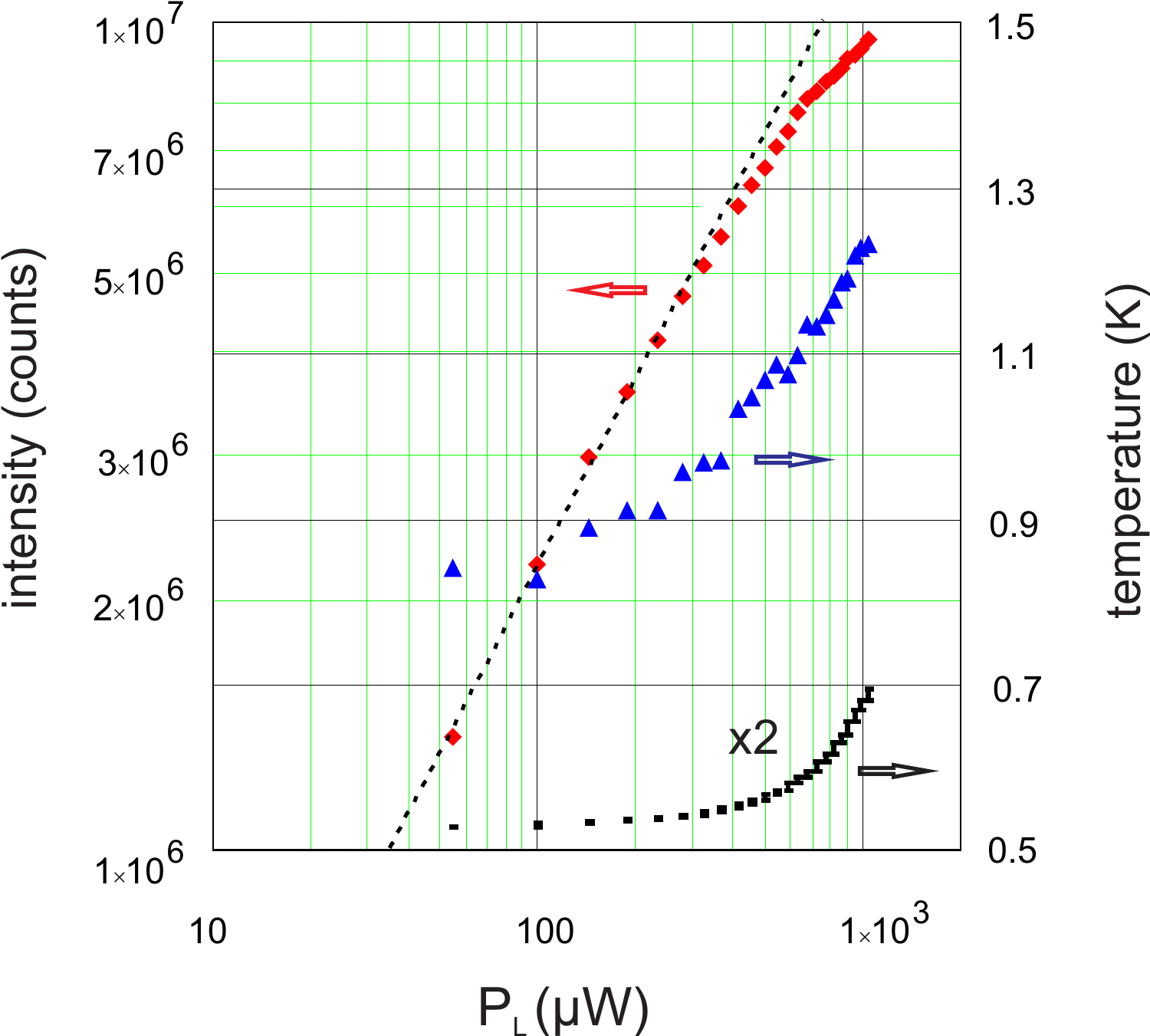}
% \end{minipage}
 \caption{Dependence of the totally integrated intensity (squares), spatial temperature (triangles), and start and final bath temperature during the measurement (black bars) on the laser power.}
 \label{fig:isothermen_all2}
\end{center}
\end{figure}

We note that the theory presented in the next section attributes such a kink in the power dependence of the total intensity of an exciton gas in a trap to the occurrence of condensation. Actually, in all of our measurements using different potential traps and under different excitation conditions such a kink shows up, provided that the Helium bath temperature is below 400 mK. Exemplarily this is shown in figures \ref{fig:isothermen_all1} and \ref{fig:isothermen_all2} for three different sets of data. In figure \ref{fig:isothermen_all1} we compare two measurements which have the same dependence of the Helium bath temperature but differ in their spatial temperatures. While at low powers both curves coincide, the curve corresponding to lower spatial temperature shows the kink at lower powers. In figure \ref{fig:isothermen_all2} we display the power dependence for the case where both the Helium bath temperature and the spatial temperature are much higher and the kink occurs at powers one order of magnitude larger than in the measurements of figure \ref{fig:isothermen_all1}. This dependence of the onset of BEC on both the bath temperature and the spatial temperature will be explained by the theory presented in the next section.  The data in figure \ref{fig:isothermen_all2} furthermore demonstrate that the origin of the kink cannot be the Auger like decay of excitons at high densities, as one might suspect. This deviation from linearity in the power dependence already shows up below the kink.   

%\clearpage

\section{Theory}\label{theory-1}

After presenting the experimental results, in this section we will provide the theoretical description, attemting to explain the main effects found in the luminescence spectra. 

So far, excitons in potential traps have been described mostly by theoretical models of non-interacting bosons. Concepts to include e.g. interaction effects have been worked out for atomic condensates. Along this line, the thermodynamics of a one-component Bose gas has been investigated extensively, see, e.g., references \cite{griffin1996,DGP99,bergeman2000,Proukakis}. First applications to excitons exist, too \cite{banyai2004}. Recently, distinct signatures of a condensate in the decay luminescence spectrum of the non-condensed excitons have been predicted using a mean-field formalism in local density approximation \cite{stolz2010}. In analogy to generalisations for multi-component atomic gases \cite{HS96,BV97,SZC00,ZYY04,CQZ05} and spinor polaritons \cite{SMK05,Ketal07}, a generalisation to a multi-component gas of interacting paraexcitons and orthoexcitons has been given in reference \cite{sobkowiak2011}.

Here, we focus on the thermodynamics of weakly interacting excitons in a trap and assume the whole exciton gas to be in thermodynamic equilibrium. This allows us to apply the theory developed in references \cite{stolz2010,sobkowiak2011}.

\subsection{Thermodynamics of trapped excitons} 
\label{sec:thermodynamics}

In order to write down the Hamiltonian of the exciton system, the interaction potential between excitons has to be modelled. The description of exciton--exciton interaction is a long-standing problem (see, e.g., \cite{SE76,CSP98,okumura2001,ZS07,SZ08,CBD08}), and so far no satisfying solution for the general case has been obtained. For our calculations, we assume a contact potential interaction $V(\mathbf{r}-\mathbf{r'})=U_{\rm 0}\delta(\mathbf{r}-\mathbf{r'})$, where the interaction strength $U_{\rm 0}$ is given by the $s$-wave scattering length $a_{\rm s}$ via $U_{\rm 0}=4\pi a_{\rm s}/M$. $M$ is the exciton mass $M=2.6 m_e$ in units of the free electron mass $m_e$ \cite{brandt2007}. This leads to the Hamiltonian in second quantisation for the grand canonical ensemble:

\begin{eqnarray}
\label{eq:Hamiltonallg}
\mathcal{H}&=& \int \mathrm{d}^3\mathbf{r} \ \psi^\dagger(\mathbf{r},t)\left( -\frac{\hbar^2 \nabla^2}{2 M} + V_{\rm ext}(\mathbf{r})-\mu \right)  \psi(\mathbf{r},t) \nonumber \\
&+&\frac{1}{2} \int \mathrm{d}^3\mathbf{r} \  U_{\rm 0} \psi^\dagger(\mathbf{r},t) \psi^\dagger(\mathbf{r},t) \psi(\mathbf{r},t) \psi(\mathbf{r},t)\;,
\end{eqnarray}
with the trap potential $V_{\rm ext}(\mathbf{r})$ and the chemical potential $\mu$.  Decomposing the field operators in the usual fashion $\psi(\mathbf{r},t)=\Phi(\mathbf{r})+\widetilde{\psi}(\mathbf{r},t)$, with the condensate wave function $\Phi(\mathbf{r})=\langle \psi(\mathbf{r},t) \rangle=\langle \psi(\mathbf{r}) \rangle$ and the operator of the thermal excitons $\widetilde{\psi}(\mathbf{r},t)$, one arrives at the Gross-Pitaevskii equation (GPE),

\begin{eqnarray}
\label{eq:GPEallg}
\hspace*{-3ex} 0 = \Bigg(\! -\frac{\hbar^2\nabla^2}{2M} +V_{\rm ext}(\mathbf{r})-\mu + U_{\rm 0} \left [ n(\mathbf{r}) + n^{\rm T}(\mathbf{r}) \right ] \!\Bigg) \Phi(\mathbf{r}) + \, U_{\rm 0} \widetilde{m}(\mathbf{r}) \Phi^*(\mathbf{r}) \,,
\end{eqnarray}
and the equations of motion for the thermal excitons,
\begin{eqnarray}
\label{eq:Bewegallg}
\mathrm{i}\hbar\frac{\partial \widetilde{\psi}(\mathbf{r},t)}{\partial t}\! &=&\! \ \Bigg(\! -\frac{\hbar^2\nabla^2}{2M} +V_{\rm ext}(\mathbf{r})-\mu + 2 U_{\rm 0}  n(\mathbf{r}) \!\Bigg) \widetilde{\psi}(\mathbf{r},t) \nonumber \\ &&+ \, U_{\rm 0} \left[\Phi^2(\mathbf{r})+\widetilde{m}(\mathbf{r})\right] \widetilde{\psi}^\dagger(\mathbf{r},t) \;,
\end{eqnarray}
with the normal and anomal averages $n^{\rm T}=\langle \widetilde{\psi}^\dagger \widetilde{\psi} \rangle$ and $\widetilde{m}=\langle \widetilde{\psi} \widetilde{\psi} \rangle$, the condensate density $n^{\rm c}=|\Phi|^2$, and the total density $n=n^{\rm c}+n^{\rm T}$.

Since the spatial extension of the potential trap is large compared to the typical length scales of the interacting exciton gas (in particular with respect to the thermal deBroglie wavelength of the excitons), we can use the local density approximation, thus treating the excitons as a locally homogeneous system. The equation of motion (\ref{eq:Bewegallg}) is solved by a Bogoliubov transformation. In the so-called Hartree-Fock-Bogoliubov-Popov (HFBP) limit ($\widetilde{m}\to 0$), the quasiparticle energy reads
\begin{eqnarray}
\label{eq:Energie}
E(\mathbf{k},\mathbf{r})\! &=&\! \sqrt{{\cal L}(\mathbf{k},\mathbf{r})^2-(U_{\rm 0}n^{\rm c}(\mathbf{r}))^2}\,, 
\end{eqnarray}
with
\begin{equation}
 {\cal L}(\mathbf{k},\mathbf{r})=\hbar^2k^2/2 M +V_{\rm ext}(\mathbf{r}) - \mu +2 U_{\rm 0} n(\mathbf{r})\,.
\end{equation}
Within these approximations, the non-condensate density $n^{\rm T}$ is given by
\begin{equation}
\label{eq:nT}
n^{\rm T}(\mathbf{r}) = \int \frac{d^3\mathbf{k}}{(2\pi)^3} \left[ \frac{{\cal L}(\mathbf{k},\mathbf{r})}{E(\mathbf{k},\mathbf{r})}\left( n_{\rm B}(E(\mathbf{k},\mathbf{r}))+\frac{1}{2} \right)-\frac{1}{2} \right] \Theta\Big (E^2(\mathbf{k},\mathbf{r})\Big ) \;.
\end{equation}
Applying the Thomas-Fermi approximation to the GPE (\ref{eq:GPEallg}), i.e., neglecting the kinetic energy term, yields the condensate density as
\begin{eqnarray}
\label{eq:nc}
n^{\rm c}(\mathbf{r}) = \frac{1}{U_{\rm 0}}\Big [ \mu-V_{\rm ext}(\mathbf{r})-2U_{\rm 0}n^{\rm T}(\mathbf{r}) \Big ] \,\Theta\Big (\mu-V_{\rm ext}(\mathbf{r})-2U_{\rm 0}n^{\rm T}(\mathbf{r}) \Big)\;.
\end{eqnarray}
Evaluating equations (\ref{eq:nT}) and (\ref{eq:nc}), $n^{\rm T}$ and $n^{\rm c}$ have to be determined self-consistently.       

\subsection{Theory of decay luminescence}

Since the optical wavelength of the emission is much smaller than the trapped exciton cloud, we apply a local approximation to the emission spectrum \cite{shi1994,haug1983} as well, which is determined by the excitonic spectral function $A(\mathbf{r},\mathbf{k},\omega)$,
\begin{eqnarray}
\label{eq:spectrum}
I(\mathbf{r},\omega) & \propto & 2 \pi \hbar |S(\mathbf{k}=0)|^2 \delta(\hbar\omega^\prime - \mu) n^{\rm c}(\mathbf{r}) \nonumber\\
&& + \sum_{\mathbf{k}\ne 0}|S(\mathbf{k})|^2n_{\rm B}(\hbar\omega^\prime - \mu)A(\mathbf{r},\mathbf{k},\hbar\omega^\prime - \mu)\,,
\end{eqnarray}
with $S(\mathbf{k})$ representing the exciton-photon coupling. The spectral function is given by the quasiparticle spectrum in (\ref{eq:Energie}):
\begin{eqnarray}
\label{eq:specfun}
A(\mathbf{r},\mathbf{k},\omega)=2\pi\hbar\left[ u^2(\mathbf{k},\mathbf{r}) \delta(\hbar\omega-E(\mathbf{k},\mathbf{r}))
-\,v^2(\mathbf{k},\mathbf{r}) \delta(\hbar\omega+E(\mathbf{k},\mathbf{r})) \right] ,
\end{eqnarray}
where $u(\mathbf{k},\mathbf{r})$ and $v(\mathbf{k},\mathbf{r})$ are the Bogoliubov amplitudes,
\begin{eqnarray}
\label{eq:bogoliubov}
u(\mathbf{k},\mathbf{r})^2 &=& \frac{1}{2} ({\cal L}(\mathbf{k},\mathbf{r})/E (\mathbf{k},\mathbf{r})+1)\,, \nonumber \\
v(\mathbf{k},\mathbf{r})^2 &=& \frac{1}{2} ({\cal L}(\mathbf{k},\mathbf{r})/E (\mathbf{k},\mathbf{r})-1)\,.
\end{eqnarray}

In Cu$_2$O, the decay of ground-state paraexcitons in the yellow series is optically forbidden. However, due to the applied strain, a mixing with the green series takes place and the decay becomes weakly allowed \cite{kreingold1975}. The paraexcitons decay directly, whereby momentum conservation requires that only excitons with the same momentum as the emitted photons are involved. This zero-phonon decay can be treated by setting $\omega^\prime=\omega-E_{\rm gX}/\hbar$ with the excitonic band gap $E_{\rm gX}$ and $|S(\mathbf{k})|^2= S_{\rm 0}^2 \delta(\mathbf{k}-\mathbf{k}_{\rm 0})$ with $\mathbf{k}_{\rm 0}$ being the wave vector of the intersection of the photon and exciton dispersions.
 
Due to the form of $S(\mathbf{k})$, a condensate of paraexcitons in Cu$_2$O (energetically in the ground state, i.e., $k=0$) should not contribute to the luminescence spectrum. However, this statement holds rigorously only for homogeneous, infinitely extended systems. Due to the finite size of the condensate, it is spread out in $\mathbf{k}$-space, and a weak contribution to the luminescence should be expected. 
Taking the condensate as a classical coherent emitter, the radiation follows from classical Maxwell equations as the Fourier transform of the polarisation \cite{bornwolf} at the wave vector of the emitted photon $\mathbf{k}_{\rm 0}$.  Since the polarisation is proportional to the condensate wave function,
the (dimensionless) strength of the contribution $S_{\rm c}$ ($|S(\mathbf{k}=0)|^2=S_{\rm c}S_{\rm 0}^2$) should be proportional to the components of the Fourier transform of the ground-state wave function at $\mathbf{k}=\mathbf{k_{\rm 0}}$. Our calculations show that $S_{\rm c}$ is of the order of $10^{-6}$ to $10^{-7}$. Therefore, we will discuss possible effects of a BEC in the spectrum considering a weakly luminescing as well as a completely dark condensate.

Furthermore, to account for the finite spectral resolution in experiments, we convolve the spectral intensity (\ref{eq:spectrum}) with a spectral response function of the shape exp$[-(\omega/\Delta)^{2}]$. Here, $\Delta$ is a measure for the spectral resolution. Using these assumptions, the spectrum reads:
\begin{eqnarray}
\label{eq:fertigdep}
I(\mathbf{r},{\omega}) &\propto& S_{\rm c} (2\pi)^3 \exp{\left[-\left(\frac{\hbar \omega ' - \mu}{\Delta}\right)^2\right]} n^{\rm c}(\mathbf{r}) \\ +&& u^2(\mathbf{k}_{\rm 0},\mathbf{r}) n_{\rm B}(E(\mathbf{k_{\rm 0}},\mathbf{r})) \exp{[-\varepsilon_{-}^2 (\omega ',\mathbf{k}_{\rm 0},\mathbf{r})]}\nonumber \\  -&&  v^2(\mathbf{k}_{\rm 0},\mathbf{r}) n_{\rm B}(-E(\mathbf{k_{\rm 0}},\mathbf{r})) \exp{[-\varepsilon_{+}^2 (\omega ',\mathbf{k}_{\rm 0},\mathbf{r})]}\, ,\nonumber 
\end{eqnarray}
with $\varepsilon_{\pm} (\omega ',\mathbf{k},\mathbf{r})\equiv(\hbar \omega ' - \mu \pm E(\mathbf{k},\mathbf{r}))/\Delta$. As explained in section \ref{sec:exp}, in the experiment, a spectrograph is used to obtain different spectra by integrating over either one or more spatial coordinates and/or $\omega$. Here, we consider the $z$-resolved luminescence spectrum $I(z,\omega)$, the $z$-profiles of the luminescence $I(z)$, the spatially integrated luminescence $I(\omega)$, and the totally integrated luminescence $I_{\rm tot}$. The $z$-resolved luminescence spectrum is obtained by imaging a small stripe of width $2\Delta y$ elongated along the $z$-direction onto the entrance slit of the spectrograph, hence integrating over the $x$- and $y$-direction. The $z$-profiles are generated by integrating $I(z,\omega)$ over the energy ($\omega$). Integrating over all spatial dimensions yields the spatially integrated luminescence $I(\omega)$. As a fourth option, one can also integrate over $\mathbf{r}$ and $\omega$, arriving at the totally integrated luminescence $I_{\rm tot}$, which only depends on the exciton number $N$ and the temperature $T$.

\subsection{Results}
\subsubsection{Luminescence spectrum}

For the calculations, we used an anharmonic potential trap fitted to the experimental of section \ref{sec:exp} but with a trap minimum of $V_{\rm 0}=1.04\ \milli\electronvolt$. The $s$-wave scattering length is chosen to be $a_{\rm s}=2.18 a_{\rm B}$ (taken from \cite{ceperley2001}, see also \cite{IMV02,SC05,SKC09}) with the excitonic Bohr radius $a_{\rm B}=0.7\ \nano\meter$.

\begin{figure}
 \begin{center}
  \includegraphics[width=0.65\textwidth]{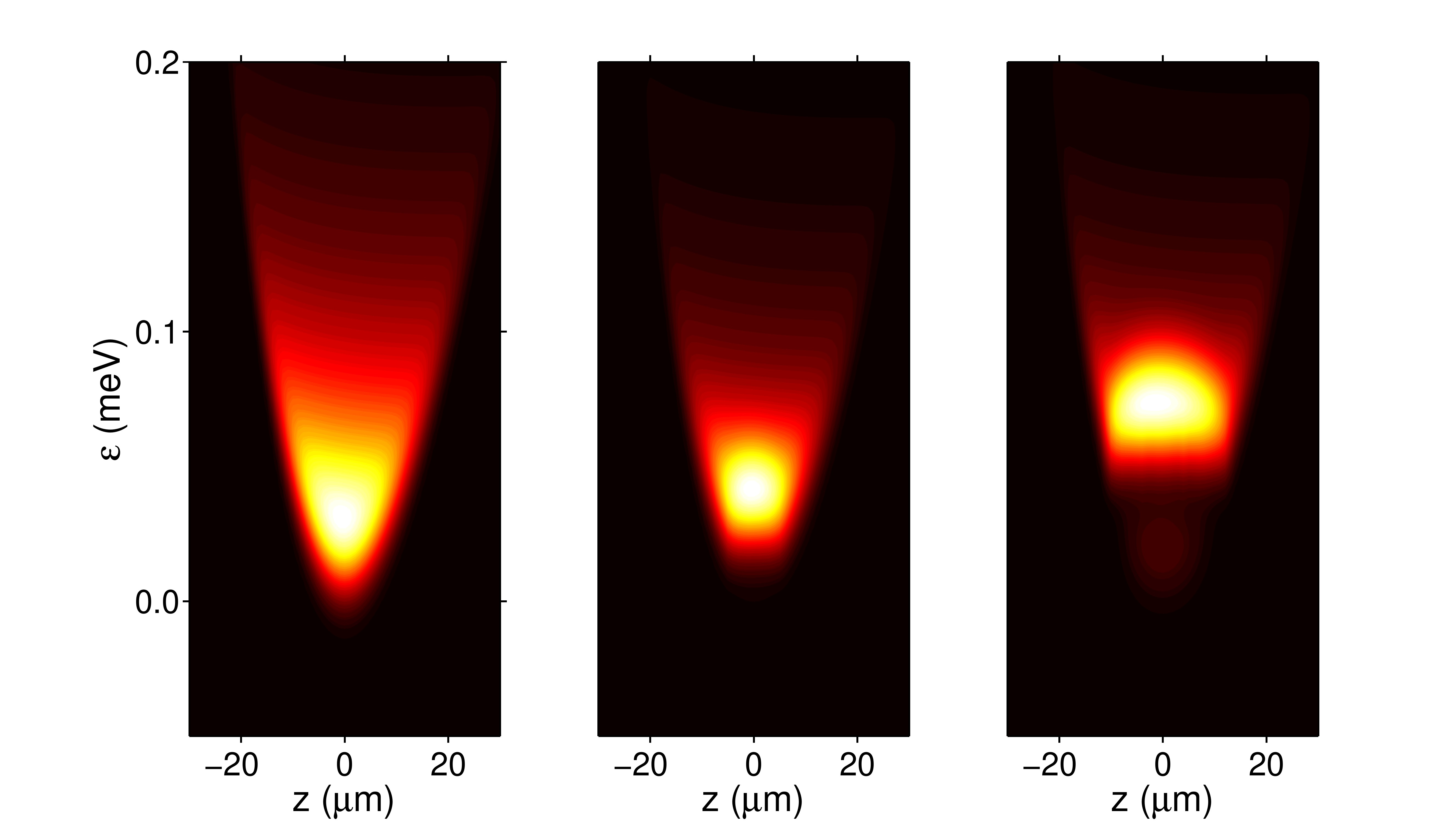}
 \end{center}
%  \begin{minipage}{8cm}
%   \includegraphics[width=\textwidth]{fig1b.pdf}
%  \end{minipage}
 
 \caption{$z$-resolved luminescence spectrum $I(z,{\omega})$ at $T=0.6\ \kelvin$ with dark condensate. The exciton numbers are $N=2.5\cdot10^7$ (left), $N=2.5\cdot10^8$ (middle) and $N=8.0\cdot10^8$ (right). Condensate fractions are $N_{\rm c}/N=0$ (left), $N_{\rm c}/N=0.05$ (middle) and $N_{\rm c}/N=0.50$ (right). Trap minimum is at $V_{\rm 0}=1.04\ \milli\electronvolt$ and $\varepsilon=\hbar\omega-E_{\rm gX}-V_{\rm 0}$.}
 \label{fig:lum}
\end{figure}

First, we revisit the flat bottom shape of the spectrum discussed in reference \cite{stolz2010}. In figure \ref{fig:lum}, we plot the luminescence spectrum for a constant temperature $T=0.6\ \kelvin$ and three different exciton numbers (left column: $N=2.5\cdot10^7$, middle column: $N=2.5\cdot10^8$, right column: $N=8.0\cdot10^8$). The condensate is expected to remain completely dark. The left image of figure \ref{fig:lum} shows a thermal spectrum exhibiting the typical nearly parabolic shape that was also found in the experimental results presented in figures \ref{fig:spectra_lowp} and \ref{fig:series_im}. Increasing the exciton number by a factor of 10 while keeping the temperature constant results in the onset of a BEC with a condensate fraction of $N_{\rm c}/N=0.05$. The shape of the spectrum is altered and develops a flat bottom at the chemical potential as reported in reference \cite{stolz2010}. Further increasing the exciton number leads to a more pronounced flat bottom and an energetic shift of the spectrum with the chemical potential. Additionally, a new contribution to the thermal spectrum below the chemical potential, arising from the $v^2$-term in equation (\ref{eq:specfun}), becomes faintly visible. These effects are linked to the occurrence of a BEC and would indicate its existence even without direct emission from the condensate.

\begin{figure}
 \begin{center}
  \includegraphics[width=1.1\textwidth]{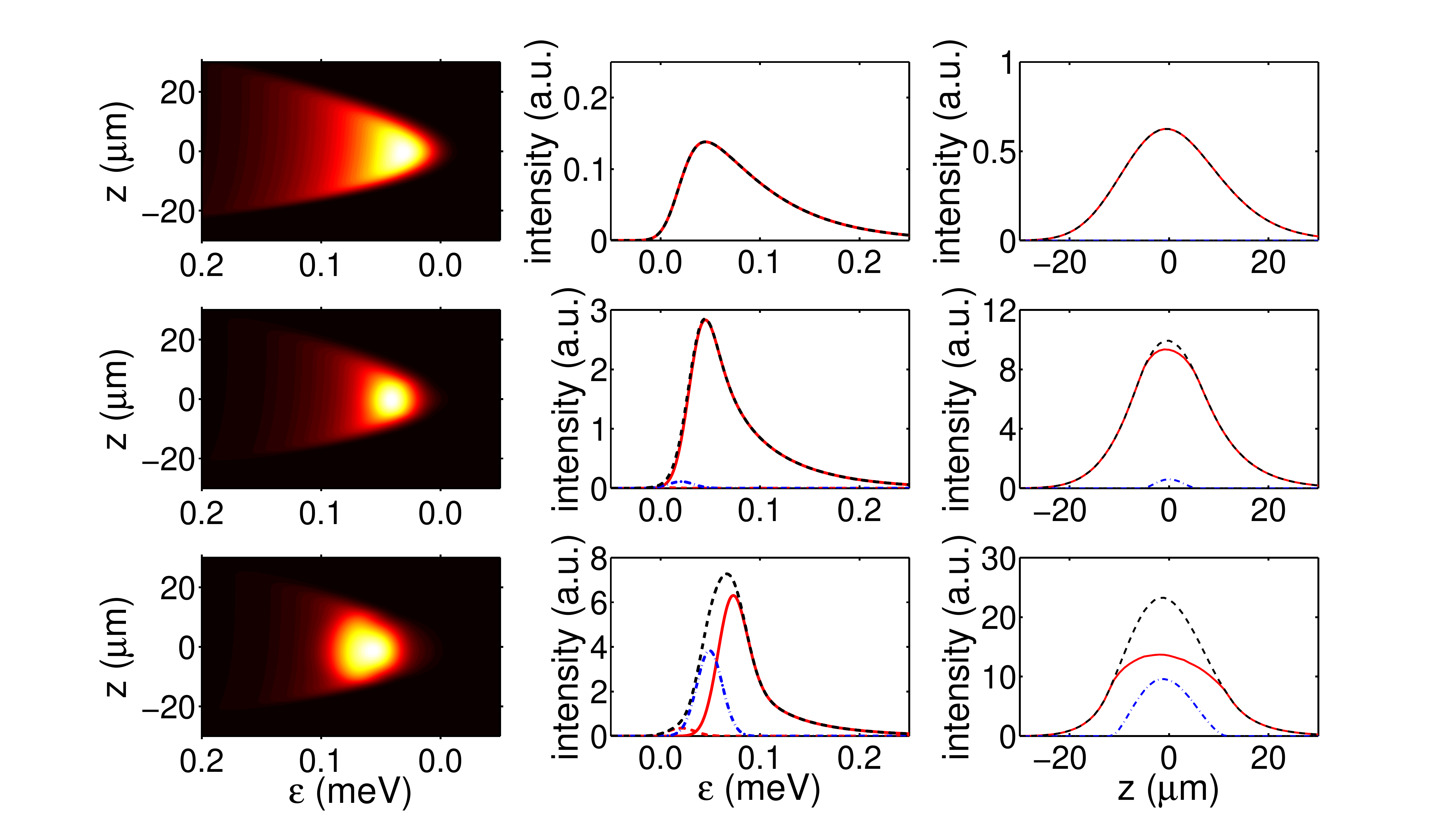}
 \end{center}
%  \begin{minipage}{8cm}
%   \includegraphics[width=\textwidth]{fig1b.pdf}
%  \end{minipage}
 
 \caption{$z$-resolved luminescence spectrum, $z$-integrated luminescence, and $z$-profiles of the luminescence with weakly luminescing condensate ($S_{\rm c}=10^{-6}/3$) for the same parameters as in figure \ref{fig:lum}. Left column: luminescence spectrum $I(z,{\omega})$. Middle column: $z$-integrated luminescence $I(\omega)$. Right column: $z$-profiles $I(z)$. Red curves (full and dashed) denote the contributions from thermal excitons, the blue dash-dotted curve gives the condensate contribution, and the total emission is shown as a dashed black line.}
 \label{fig:lum1}
\end{figure}

Comparing these predictions with the experimental findings, e.g., figures \ref{fig:spectra_lowp} and \ref{fig:series_im}, one has to conclude that the signatures were not observed in the experiment.
However, if the condensate exhibits a very weak luminescence, the changes in the spectrum predicted from theory are much more subtle. This can be seen in the left column of figure \ref{fig:lum1} ($S_{\rm c}=10^{-6}/3$), cf.\ again figures \ref{fig:spectra_lowp} and \ref{fig:series_im}. In contrast to the case of a dark condensate, there is no drastic qualitative change in the spectra from top to bottom. Also, the shift on the energy axis appears to be smaller and could well be an interaction effect of thermal excitons. Therefore, the contribution from the condensate hides the flat bottom as well as the shift on the energy axis, without being a clearly visible delta-shaped peak, as expected for a fully contributing condensate.
This explains why a conclusive detection of a condensate via spectral signatures requires a very careful analysis of the experiments.

The spectra in the left column of figure \ref{fig:lum1} consist of different contributions. To illustrate this, we plot the results for the spatially integrated luminescence $I(\omega)$ in the middle column of figure \ref{fig:lum1}, cf.\ the measurements shown in figures \ref{fig:spectra_lowp} and \ref{fig:powerspectra}. The upper panel shows a spectrum in the non-condensed case at low densities. Its shape is given by a Bose distribution convolved with the spectral resolution of the spectrometer. The middle panel shows a spectrum at higher densities, where interaction effects are already important and a very small condensate contribution already occurs, nearly invisible in the total curve. Compared to the previous case, the peak becomes narrower and shifts to higher energies, while the tail remains qualitatively the same. The bottom graph shows the case with a condensate fraction of $0.5$ including a distinct condensate contribution (dashed-dotted blue) which has a Gaussian shape and is centred at the chemical potential. The contribution from the thermal excitons consists of the $u^2$ part of the spectral function (solid red) and $v^2$ part of the spectral function (dashed red), compare equation (\ref{eq:fertigdep}). The black line represents the sum over all contributions. In contrast to the very weakly condensed case (middle), the contribution related to the $v^2$ term appears below the chemical potential (dashed red). Except for this additional contribution, the total emission is only slightly shifted towards higher energies and has a higher maximum compared to the middle graph in figure \ref{fig:lum1}. The width of all the Gaussian like peaks in figure \ref{fig:lum1} is directly given by the spectral resolution $\Delta=18\,\micro$eV. 

\subsubsection{Spatially resolved luminescence}

The results for the spatially resolved luminescence $I(z)$ are presented in the right column of figure \ref{fig:lum1} and should be compared to the experimental results shown in figures \ref{fig:spectra_lowp}, \ref{fig:spatial_im}, \ref{fig:comp_820_1617}, and \ref{fig:comp_z}.

Without any contribution from the condensate, one would expect the solid red line. The non-condensed case follows a Gaussian shape (top graph), while the onset of the BEC leads to a deformation in the form of a plateau (middle and bottom graph). However, taking a weakly luminescing condensate (dashed blue) into account, the total emission (black dashed) looks Gaussian like again, masking the signature of the condensate (middle and bottom graph). 

\subsubsection{Totally integrated luminescence}
\label{sec:inegratedlum}
\begin{figure}[h]
\begin{center}
\includegraphics[width=0.85\linewidth]{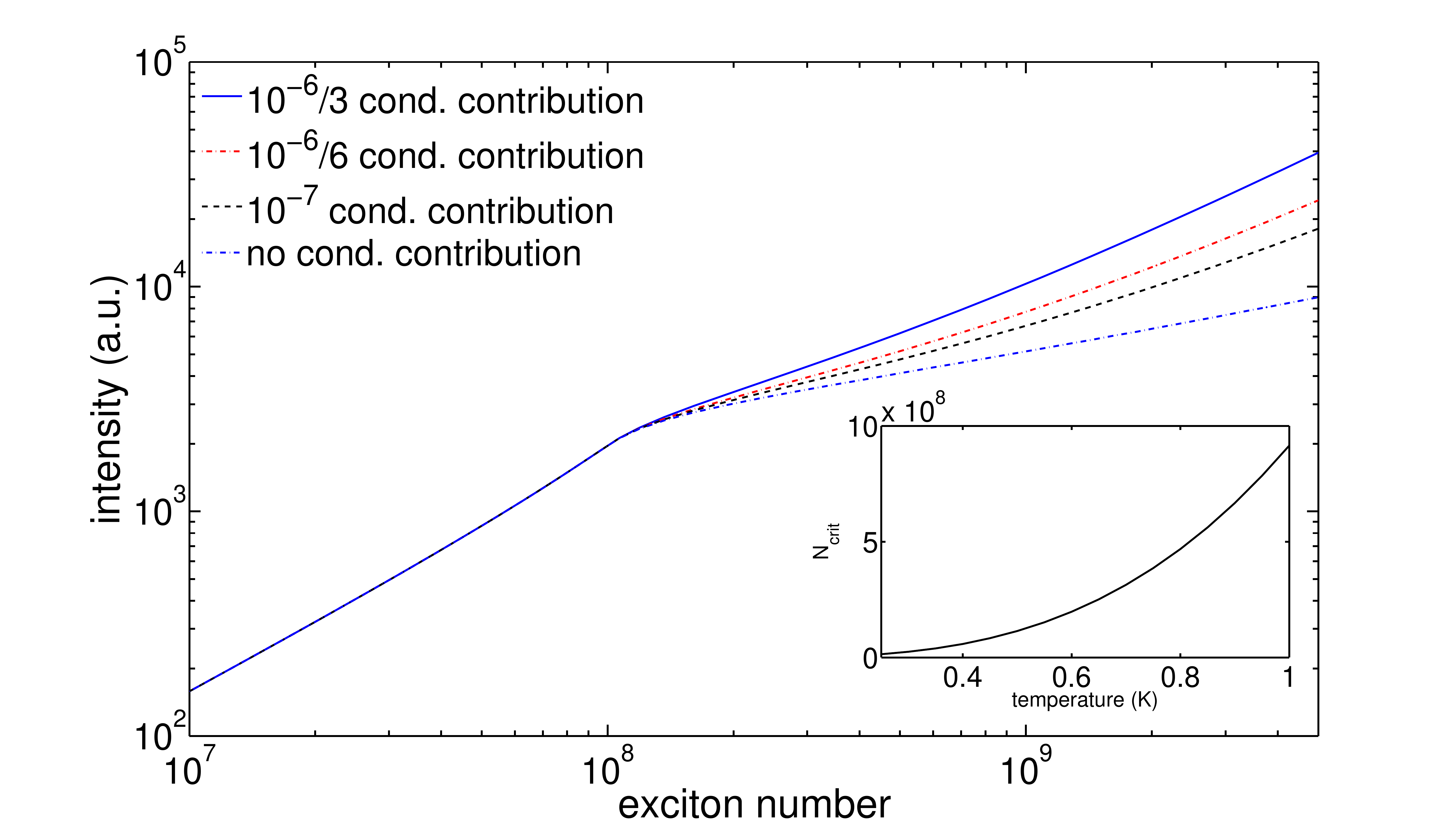}
\caption{Totally integrated luminescence $I_{\rm tot}$ for $T=0.5\ \kelvin$ and different fractions of condensate contribution. Change of $N_{\rm crit}$ (position of kink) with temperature shown in the inset.}
\label{fig:isotherme}
\end{center}
\end{figure}

Integrating over all variables ($\mathbf{r}$ and $\omega$) yields the totally integrated luminescence $I_{\rm tot}$, which had been considered in the experiments in figures \ref{fig:isotherm2510}, \ref{fig:isothermen_all1}, and \ref{fig:isothermen_all2}. In figure \ref{fig:isotherme}, we show the numerical results for $I_{\rm tot}$ as a function of the exciton number $N$ for a fixed temperature of $T=0.5\ \kelvin$. Remember that $N$ is connected to the excitation laser power $P_{\rm L}$, cf.\ the results of the rate model, figure \ref{fig:rate_model}. For small $N$, the totally integrated luminescence increases linearly with the exciton number until a critical value is reached. At the critical exciton number, the curve has a kink and continues with a weaker slope afterwards. The additional contribution from a weakly luminescing condensate does not alter the behaviour qualitatively in this case, as it can be seen from the other curves in figure \ref{fig:isotherme}. As shown in the inset of figure \ref{fig:isotherme}, the critical exciton number shifts with the temperature approximately as expected with $T^3$, equation (\ref{eq:Ncid}). Obviously, here the interaction has no drastic effect on the behaviour known from an ideal gas.  

\begin{figure}[h]
\begin{center}
\includegraphics[width=0.85\linewidth]{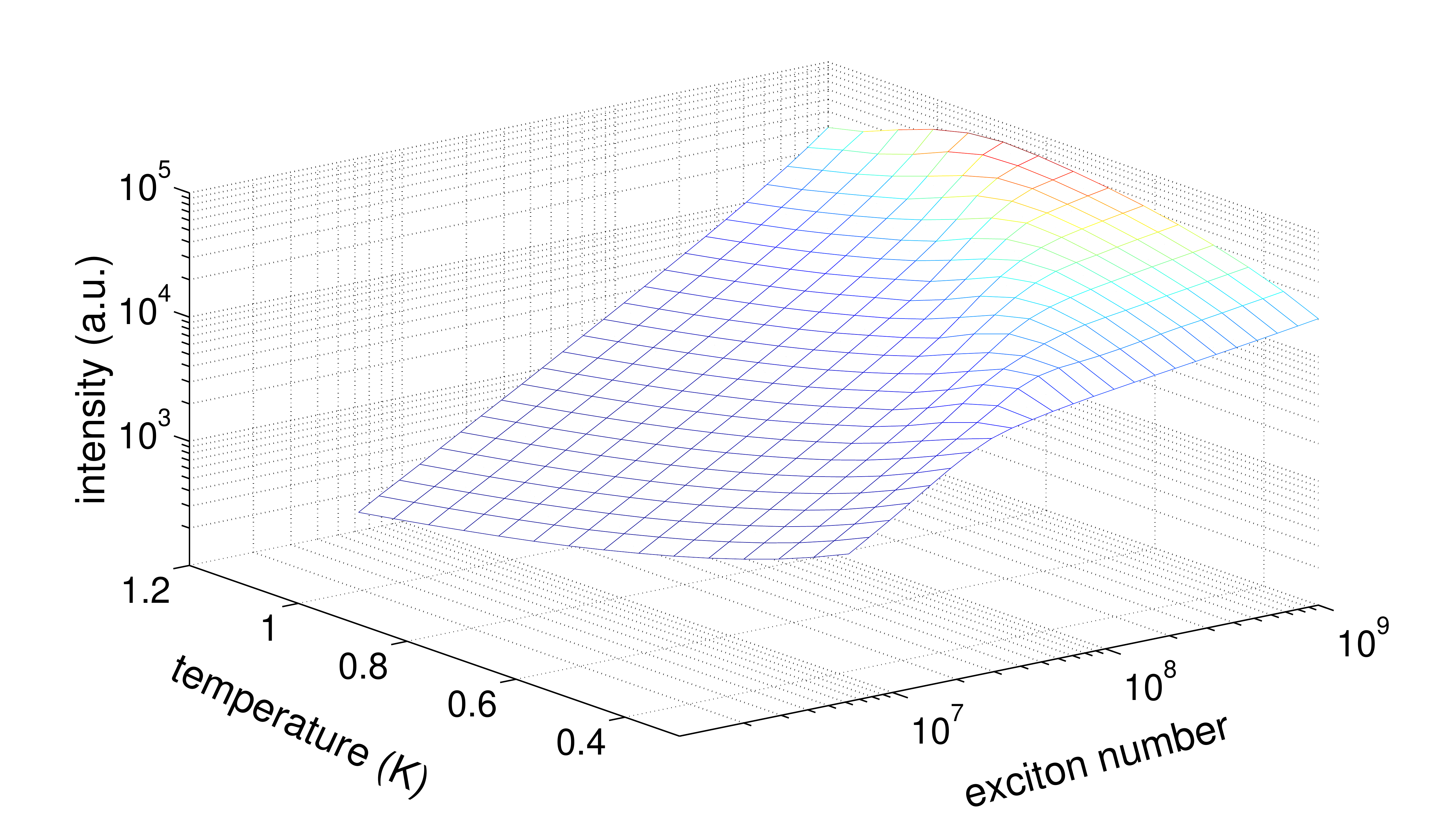}
\caption{Totally integrated luminescence $I$ for a range of different temperatures and exciton numbers in the same potential trap as in figure \ref{fig:lum}.}
\label{fig:INTebene}
\end{center}
\end{figure}

However, in the experimental situation, it is merely impossible to measure over orders of magnitude of the particle number without changing the exciton temperature. The temperature will rise as a result of the energy introduced into the system by the laser. Therefore, experimental points may lie anywhere in the $N$--$T$ plane. The theoretical results for the totally integrated luminescence as a function of exciton number and temperature are shown in figure \ref{fig:INTebene}. Depending on how fast the temperature increases with exciton number, it is very well possible to never cross the phase boundary and observe the kink in the totally integrated luminescence.  

\subsection{Discussion}

In the previous section, we presented different spectral signatures of a non-emitting condensate in the thermal emission of the excitons in global thermal equilibrium. The five main signatures are: (i) the formation of a flat bottom in the luminescence spectrum $I(z,\omega)$, (ii) the shift on the energy axis of the spectrum $I(z,\omega)$, (iii) a deviation of the $z$-profiles $I(z)$ from the Gaussian shape, (iv) the appearance of the $v^2$ term in the spatially integrated luminescence $I(\omega)$, (v) the kink in the totally integrated luminescence $I_{\rm tot}$ at $N_{\rm crit}$.

If the contribution from the condensate is much weaker than what we estimated (i.e., $S_{\rm c}$ of the order of $10^{-6}...10^{-7}$), signature (i) should be experimentally visible. This is neither the case in the current experiment nor have other experiments seen this effect. If the contribution is much stronger though, one should expect the delta-shaped peak as predicted by the standard theory \cite{blatt1962,haug1983,shi1994}. This substantiates our estimation for $S_{\rm c}$.
 
However, if the condensate contribution is comparable to that of the thermal excitons, the flat bottom in (i) can be masked as shown in figure \ref{fig:lum1}. The changes in the shape are very subtle and probably not detectable in an actual experiment. The shift on the energy axis (ii) is also altered by the condensate contribution. Additionally, there are numerous other effects that can change the energetic position of the spectrum, e.g., a background plasma of electrons and holes \cite{semkat2009,manzke2010}. Therefore, this shift might not be a good indicator for the onset of the BEC. The deformation of the $z$-profiles (iii) would also be masked by the condensate contribution. Here, the latter one complements the thermal luminescence, again resulting in a Gaussian profile. Contrarily, in the measured profiles (figure \ref{fig:comp_z}), the condensate adds a contribution to the thermal Gaussian profile. We will revisit this point later.

Without a condensate, the spectra can be fitted with a renormalised Bose distribution convolved with the spectral resolution of the spectrometer. With a condensate, the $v^2$ peak (iv) causes a characteristic deformation of the low energy flank of the spatially integrated luminescence. For a dark condensate, however, it would be a free standing peak separated from the rest of the spectrum. Both cases should be detectable in an experiment though. However, this signature (iv) as well as the energetic shift (ii) are very subtle and are probably masked by noise, compare figures \ref{fig:spectra_lowp} and \ref{fig:series_im}.

The most promising signature of condensation would be the kink in the totally integrated luminescence (v), as it is not altered qualitatively by the condensate contribution. In this case it only changes the slope after the kink, but does not hide the kink itself. Although in the figures \ref{fig:isotherm2510}-\ref{fig:isothermen_all2} no isotherms are plotted -- the exciton temperature rises with excitation power -- there is obviously a kink in the measured integrated intensity. Looking at figure \ref{fig:INTebene}, a path on the surface will exhibit a kink even with increasing temperature, if the ridge is crossed. Therefore, we can relate the kink in the experimental figures, at least qualitatively, to the occurrence of a condensate.

\subsection{Excitons in local equilibrium}

As stated above, the theory derived in section \ref{theory-1} assumes that the exciton gas in the trap is in global thermodynamic equilibrium. The experimental results presented in the previous section show, however, that this is obviously not the case and that the exciton temperature is the key quantity to detect deviations from global equilibrium.
The conclusions are that (i) the \textit{spectral} (or \textit{spatial}) temperature is not equal to the exciton temperature and (ii) the assumption of global equilibrium, therefore, must be wrong, cf.\ the discussion in section \ref{sec:lowp}.

In contrast to the latter statement, the luminescence spectra calculated by the equilibrium theory reproduce the measured spectra qualitatively quite well, cf.\ figures \ref{fig:spectra_lowp} and \ref{fig:lum1}. Thus it seems to be reasonable to abandon the global equilibrium assumption, but to keep local equilibrium. In this case, the thermally excited excitons have still an equilibrium (Bose) distribution, but with spatially varying temperature and chemical potential. The space dependence of the latter quantities is in principle unknown, but from the experiment we can conclude that the temperature should not vary over the dimension of the trap, see section \ref{sec:lowp}. The only criterion for the spatial dependence $\mu(\mathbf{r})$ is that the spectra must be reproduced. Therefore, we demand that the spatially integrated spectrum $I(\omega)$ (at least its high-energy tail) follows a Bose distribution with the spatial temperature and some formal constant chemical potential ${\tilde\mu}$ which is used to fix the particle number. This leads to
\begin{equation}\label{my_r}
\mu(\mathbf{r})=\left(1-\frac{T_{\rm X}}{T_{\rm s}}\right)\left(V_{\rm ext}(\mathbf{r})+E_{\rm 0}+2U_{\rm 0}n^{\rm T}(\mathbf{r})\right)+\frac{T_{\rm X}}{T_{\rm s}}{\tilde\mu}\,,
\end{equation}
which allows to directly use the theory of section \ref{sec:thermodynamics} also in this nonequilibrium situation.
\begin{figure}[h]
 \begin{center}
  \includegraphics[width=0.9\textwidth]{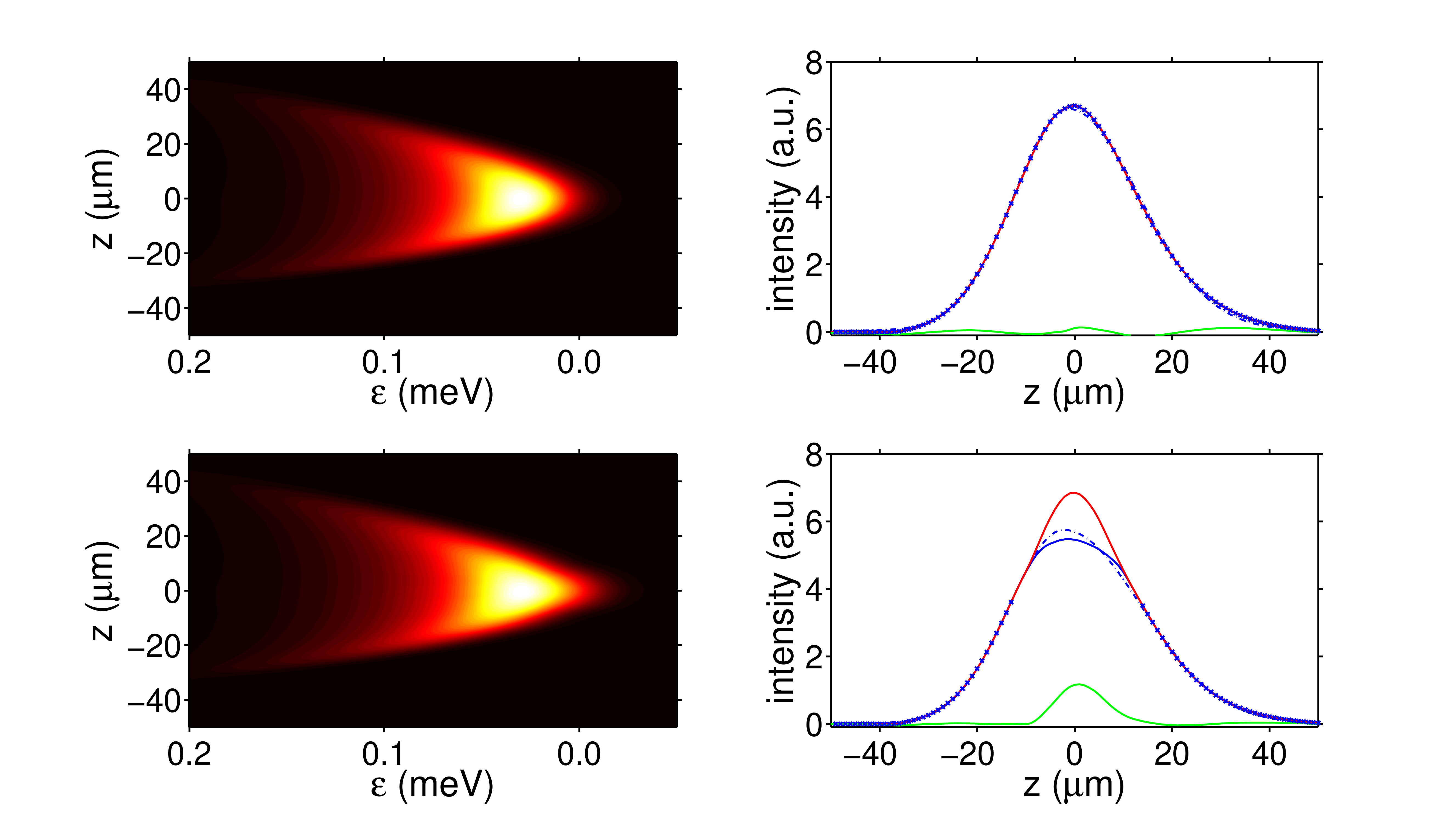}
  \end{center}

 \caption{Spatially resolved luminescence spectrum (left) and $z$-profiles of the intensity (right) for $N=10^8$ and $T_{\rm s}=0.6$ K. Upper row: exciton temperature $T_{\rm X}=0.25$ K (uncondensed case), lower row: $T_{\rm X}=0.15$ K (condensed case, $\eta_c=0.2$). Solid red lines: total profiles (thermal + condensate), solid blue: only thermal contribution, dash-dotted blue: Gaussian fit of the total profiles, blue crosses: data points used for the fit, solid green: difference between total profile and fit. The ratio between integrated intensities of the condensate and of the total band is $\eta_{\rm lum}=0.07$}.
 \label{fig:lumvorundnachkond}
\end{figure}

In what follows, we apply the local equilibrium theory to two typical experimental situations.
Thereby we fix the total particle number $N=10^8$ and the spatial temperature $T_{\rm s}=0.6$ K and vary the exciton temperature. The condensate contributes to the luminescence with $S_{\rm c}=4\cdot10^{-7}$. The spatially resolved spectra for both cases are displayed in figure \ref{fig:lumvorundnachkond}, left column. As expected, in the uncondensed case ($T_{\rm X}=0.25$ K, upper row), the spectrum just follows the external potential. The condensed case ($T_{\rm X}=0.15$ K, lower row)  with a condensate fraction $\eta_c=0.2$ looks qualitatively not very different. In particular, due to the spatial variation of the chemical potential according to equation (\ref{my_r}), the flat bottom of the spectrum (cf.\ figure \ref{fig:lum}) disappears.

In the condensed case ($T_{\rm X}=0.15$ K), the $z$-profile of the luminescence exhibits a clearly non-Gaussian shape (figure \ref{fig:lumvorundnachkond}, right column, lower panel, red line). For the not much higher temperature of $T_{\rm X}=0.25$ K (upper panel), the shape is approximately Gaussian. Therefore, we fit the thermal component of the $z$-profile for the condensed case by an asymmetric (due to the potential asymmetry) Gaussian. In order to exclude the condensate contribution, we omit in the fitting procedure the data points in the centre of the trap where the condensate is situated. The size of the excluded area is determined by minimising the error of the fit.
The fitting result is given by the dash-dotted blue line in the lower right figure. It basically follows the thermal profile component. The small deviations from the Gaussian shape are caused by the potential anharmonicity and by the renormalisation of the potential due to the interparticle interaction. Thus, the procedure to extract the thermal component from the experimental $z$-profiles shown in figure \ref{fig:comp_z}, where a Gaussian fit has been applied, too, seems to be justified.
The difference between the total $z$-profiles and the Gaussian fits is given by the green lines. It basically reflects the condensate contribution. For the condensed case we obtain a luminescence fraction $\eta_{\rm lum}=0.07$ which would set $f_{\rm lum}=0.25$.

The two cases depicted in figure \ref{fig:lumvorundnachkond} can be compared to the measured $z$-profiles shown in figure \ref{fig:comp_z}. The ``uncondensed case'' (upper row in figure \ref{fig:lumvorundnachkond}) obviously corresponds, e.g., to measurements 1 and 21, where the difference between data and fit is just noise. On the other hand, the ``condensed case'' (lower row in figure \ref{fig:lumvorundnachkond}) finds its counterparts, e.g., in measurements 3 and 17. In each case, a striking qualitative agreement is found. This corroborates our explanation of the experimental findings with the occurrence of a excitonic BEC substantially.

\begin{figure}[b]
 \begin{center}
  \includegraphics[width=0.65\textwidth,angle=-90]{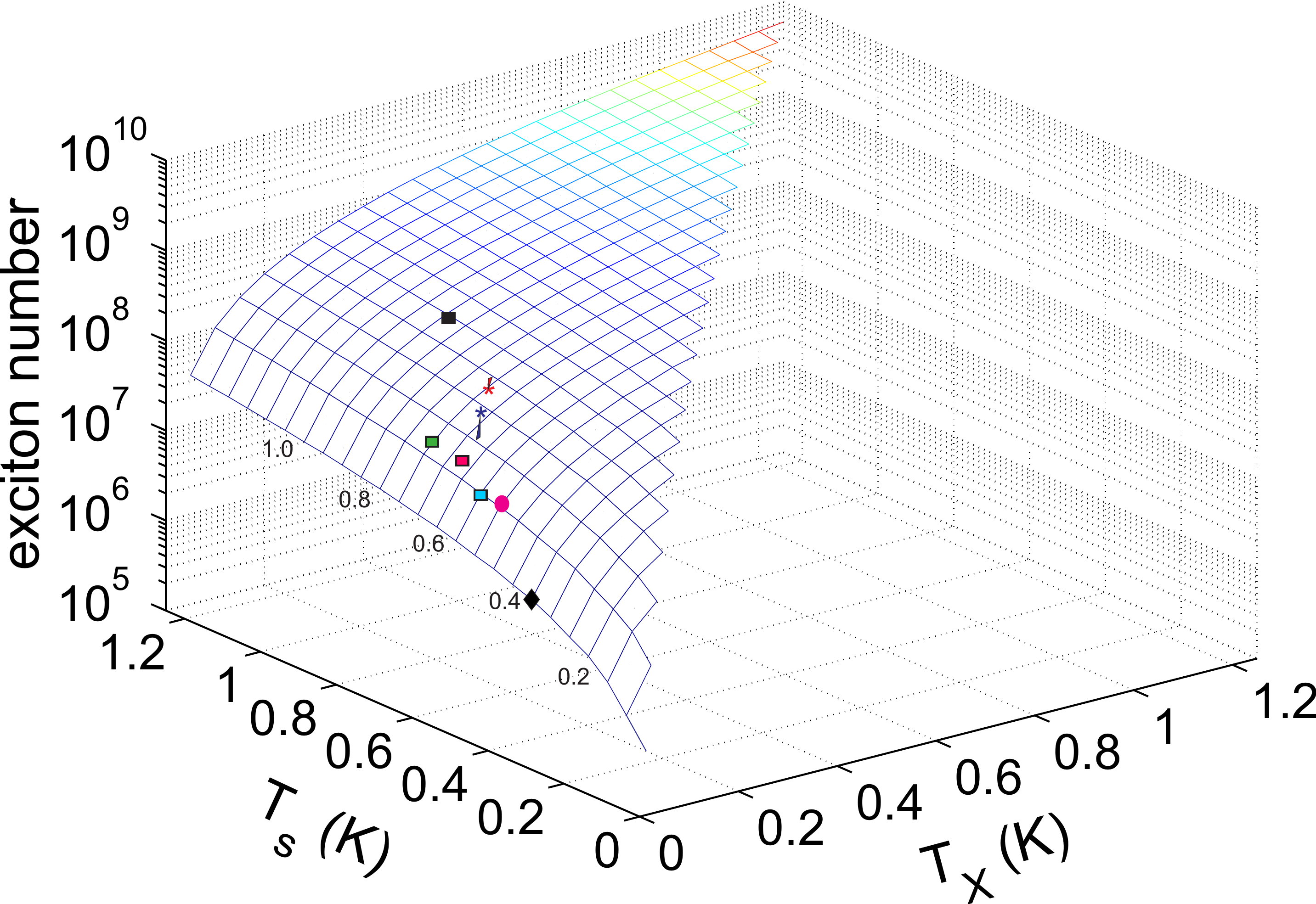}
 \end{center}
 \caption{Critical particle number $N_{\rm crit}$ for Bose-Einstein condensation vs.\ exciton temperature $T_{\rm X}$ and spatial temperature $T_{\rm s}$. The diagonal $T_{\rm X}=T_{\rm s}$ corresponds to global equilibrium; $T_{\rm X}>T_{\rm s}$ is, for excitation outside the trap, physically meaningless. The blue coloured rectangle corresponds to the critical power of $30\,\micro$W for the data shown in figure \ref{fig:isothermen_all1} (blue symbols), the red coloured rectangle corresponds to the critical power of $60\,\micro$W for the data shown in figure \ref{fig:isothermen_all1} (red symbols), the black coloured rectangle corresponds to the critical power of $600\,\micro$W for the data shown in figure \ref{fig:isothermen_all2}, and the green coloured rectangle corresponds to the critical power of $78\,\micro$W (data not shown). The magenta coloured dot corresponds to the critical power of $40\,\micro$W from the measurements shown in figure \ref{fig:isotherm2510}. The stars denote the conditions of measurements 17 (blue) and 18 (red), the black diamond denotes the critical number for measurement 3, all of figure \ref{fig:series_im}.}
 \label{fig:nctstx}
\end{figure}

Figure \ref{fig:nctstx} shows the critical particle number for Bose-Einstein condensation in dependence on exciton temperature $T_{\rm X}$ and spatial temperature $T_{\rm s}$. The global equilibrium case is represented by the diagonal $T_{\rm X}=T_{\rm s}$. If the excitons are excited outside the trap, their spatial profile will be at first always broader than in equilibrium. Therefore, $T_{\rm s}<T_{\rm X}$ can be ruled out in the current experiment.
Obviously, for a given exciton temperature, the critical number is smallest for global thermal equilibrium and increases with increasing spectral temperature. Even at low powers, i.e., for quite low exciton numbers, a condensate is possible if the exciton temperature is close to the He-bath temperature, cf.\ the situation shown in figure \ref{fig:comp_z}, measurement 3, where the spatial temperature is 0.35 K and the exciton number is $4\cdot 10^6$ which just corresponds to the critical number at $T_{\rm X}=0.04$ K.

With the help of this figure, we can systemise the experimentally observed power dependence of the totally integrated intensity shown in figures \ref{fig:isotherm2510}--\ref{fig:isothermen_all2}. This is done by looking at which exciton temperature $T_{\rm X}$ the critical number (corresponding to the critical power and the respective spatial temperature) crosses the critical surface. These points are plotted as coloured rectangles in figure \ref{fig:nctstx}.
The values for $T_{\rm X}$ give a very systematic picture. With increasing critical power, both the spatial and the exciton temperature rise. Thereby, $T_{\rm X}$ is always larger than the bath temperature, which is consistent with the expected crystal heating, compare \ref{app:thermal}.

Figure \ref{fig:nctstx} also compares the data for the measurements 17 (blue star) and 18 (red star) of figures \ref{fig:series_im} and \ref{fig:comp_z} showing that the conditions in measurement 17 are such that we are well above $N_{\rm crit}$ while for measurement 18 we are just touching the critical surface.

\section{Conclusions and Outlook}\label{sec:concl}

We have presented a series of experiments that investigates the luminescence from paraexcitons in cuprous oxide after cw-excitation. The excitons have been confined in a potential trap with the host crystal immersed in liquid helium at temperatures down to 38 mK.  The spatially resolved luminescence spectra do not show notable variations when varying the laser excitation power from below 1 $\micro$W to above 1 mW. However, we observed a number of characteristic changes in the spatial profiles of the luminescence and in the totally integrated intensity: In the spatial profiles, a new component in the centre of the trap occurs at intermediate laser powers which cannot be fitted by a thermal Gaussian distribution. This component vanishes at very low power and at very high power levels. At the mentioned intermediate power levels, we see deviations from the linear dependence of the totally integrated intensity on the laser power: Most notably a well defined kink appears in the slope of the curves. From the linear dependence of the total intensity on laser power, we conclude that two-body decay processes of the excitons do not play any significant role. This is in agreement with previous measurements under pulsed excitation in the same samples \cite{schwartz2011}.

With increasing laser power, we observe a slight increase in the effective temperature that is inferred from the high-energy tail of the luminescence spectra from 0.35 K up to a maximum of 1 K. This spectral temperature does only reflect the spatial distribution of excitons in the trap and is not identical to the local temperature which determines the energy distribution of the excitons at each point in the trap. The local temperature will be determined by exciton relaxation processes and is expected to go down almost to the bath temperature at low excitation power.

Recently, measurements \cite{gonokami2011} reported a strong heating of the exciton gas at high pump powers and claimed that this effect originates from a relaxation explosion of excitons when a transition into a Bose-Einstein condensate takes place. We observe a similar heating under high excitation powers but can definitely rule out the existence of a BEC in this high power range.

In order to understand the observed features of the luminescence spectra, we theoretically analysed the thermodynamics and the luminescence properties of excitons in a potential trap. Thereby the excitons are described as an interacting Bose gas in the framework of a Hartree-Fock-Bogoliubov-Popov approach. Already under the assumption of global thermal equilibrium, this theoretical approach suggests that the observed kink in the totally integrated intensity signals the transition into a Bose-Einstein condensate of trapped paraexcitons. Taking into account the specific non-equilibrium situation in the trap, the theory also consistently describes the characteristic changes in the spatial profiles of the luminescence: While the theory predicts almost no changes in the spatially resolved spectra, it allows to identify the additional component in the spatial profiles as due to a weakly luminescing condensate.

To conclude, we have presented strong evidence, both from experiment and theory, that at ultracold temperatures in the range of 100 mK, paraexcitons in Cu$_2$O undergo a transition into a Bose-Einstein condensate.

Nevertheless, further experimental investigations are necessary to prove the existence of a BEC of excitons in cuprous oxide beyond any doubt. For example, direct measurements of the lattice temperature and of the local exciton temperature will provide a better understanding of the thermodynamics. Thereby, it might be intriguing to apply spatially resolved Brillouin scattering since the energies of the phonons involved are comparable to the thermal energy. The local exciton temperature would be detectable via infrared absorption of the $1S-2P$ transition \cite{gonokami2005,klingshirn2005}. 
The most important aspect, however, must be the direct proof of the macroscopic coherence of the condensate by interferometric methods or by means of intensity correlation measurements, both being under way.

From the side of the theory, one can try to improve the exciton density calculations by using more advanced approximations, e.g., by solving the Gross-Pitaevskii equation exactly instead of using the Thomas-Fermi approximation. It also seems to be necessary to improve the luminescence theory in order to account for the inhomogeneity of the system more rigorously, e.g., following the ideas of reference \cite{Henneberger}. Moreover, one should also include the spectral broadening due to exciton-exciton interaction. This would require the inclusion of higher order correlations beyond the Hartree-Fock-Bogoliubov-Popov approximation by calculating the densities and the spectral function on the level of the Beliaev approximation \cite{hohenberg1965}.
%Both steps are important towards a rigorous luminescence theory for an excitonic BEC.
% So far, for the coupling matrix element $S(\mathbf{k})$ we have assumed $|S(\mathbf{k})|^2= S_{\rm 0} \delta(\mathbf{k}-\mathbf{k}_{\rm 0})$. 
Recently, we have shown that, already for the thermal excitons, a more realistic description requires to take the lifetime broadening of the exciton states into account \cite{schwartz2011}, which relaxes the strict wave vector conservation in (\ref{eq:spectrum}).
% If $\hbar \gamma$ is the lifetime broadening of the exciton state at $E_{\rm 0}=\hbar^2k_{\rm 0}^2/2M_P$, then %%@
% the uncertainty in the wave vector is simply given by $\Delta k = \sqrt{M_P/2 E_{\rm 0}}\cdot \gamma$ and the density of the states contributing to the %%@
% luminescence is just 
% \be
% D'_{ZP}=\frac{1}{(2\pi)^3}\int d\Omega k_{\rm 0}^2 \Delta k = \frac{k_{\rm 0} M_P}{2 \hbar \pi^2} \gamma= D_{\rm 0} \gamma
% \ee
% with which (\ref{eq:fertigdep}) has to be multiplied to get the proper intensity. For the exciton damping $\gamma$, two contributions are important %%@
% \cite{brandt2007}: Scattering by acoustical phonons, which leads to a contribution $\gamma_{\rm 0}$ that depends only on the temperature of the surrounding %%@
% phonon bath, and exciton-exciton scattering. Inclusion of the latter one would go beyond the present theoretical level of the HFBP approach leading to the so-called Beliaev approximation. To avoid that, the influence of collisions might in a first, empirical step be accounted for by a damping factor $\gamma_{XX}$ which can be assumed to depend on the local thermal exciton density $n^{\rm T}(\mathbf{r})$, %%@
% the mean exciton velocity $\overline{v}=\sqrt{18k_B T/(\pi M_P)}$, and the scattering cross section $\sigma_{XX}= 8 \pi a_S^2$ as \cite{brandt2007}
% \be
% \gamma_{XX}= \sigma_{XX} n^{\rm T}(\mathbf{r}) \overline{v}(T) \; .
% \ee

Quite recent results by Naka \textit{et al.} \cite{naka2012} show that the Auger-decay produces a significant number of free electrons and holes which are also captured by the trap (see figure \ref{fig:rate_model}). This should lead to a shift and a broadening of the exciton states by the surrounding electron-hole plasma \cite{NaSt01,semkat2009,manzke2010,semkat2010}. An inclusion of this ``plasma damping'' requires to include collisions of the excitons with charged fermions in the theory. This is clearly beyond the scope of the theory of weakly interacting bosons.

% Beside the signatures presented in this paper, one can also imagine unique signatures of a weakly luminescing condensate. Depending on the actual width of the condensate peak and the strength of its emission, it might be possible to observe something like a double peak structure in the spatially integrated luminescence. One peak would be formed by the $u^2$ term from the thermal spectrum and the other one by the condensate peak.

\ack We thank Dietmar Fr\"ohlich (Dortmund) for supplying the sample. Furthermore, we thank him, G\"unter Manzke and Wolf-Dietrich Kraeft (Rostock), Manfred Bayer, Jan Brandt, and Christian Sandfort (Dortmund), and Andreas Alvermann (Greifswald) for many helpful discussions and critical comments. Our special thanks go to Herwig Ott (Kaiserslautern) for lending us the CR599 laser system. We acknowledge the %%@
support by the Deutsche Forschungsgemeinschaft (Collaborative Research Center SFB 652 ``Starke Korrelationen im Strahlungsfeld'') and by JSPS KAKENHI (Grant No. 21740227).

\appendix
\section{Strain Hamiltonian for electron-hole states}\label{app:strain_eh}
For the derivation of the energy shifts of electron-hole pairs we consider that the top $\Gamma_7^+$ valence band states can be written as \cite{waters}
\bea
\Psi^7_{+1/2}&=&- \frac{1}{\sqrt{6}}\left[ (Y^{-2}_2 - Y^{2}_2)\alpha_v + 2 Y^{-1}_2 \beta_v \right]\,, \\
\Psi^7_{-1/2}&=&+ \frac{1}{\sqrt{6}}\left[ (Y^{-2}_2 - Y^{2}_2)\beta_v - 2 Y^{1}_2 \alpha_v \right]\,.
\eea
Here $Y^{m}_2$ denote the spherical harmonics and $\alpha_{\rm c,v}, \beta_{\rm c,v}$ the electron spin functions of valence and conduction band.
By inspection of table VI of reference \cite{waters} we immediately see that the paraexciton state 
\be
\Phi_{12} = \Phi_{\rm YS} \frac{1}{\sqrt{2}}\left (\Psi^7_{-1/2}\alpha_{\rm c} + \Psi^7_{+1/2} \beta_{\rm c} \right)
\ee
with $\Phi_{\rm YS}$ being the envelope function of the yellow $1S$ state. We immediately see that the paraexciton state has the same behaviour as the $\Gamma_7^+$ band states. In all matrix elements of the strain Hamiltonian, all terms containing the electron-hole exchange are missing. Since the interaction with the $\Gamma_8^+$ states does not change, the electron-hole pairs show the same behaviour under strain, i.e., they perceive an effective trapping potential similar to the paraexcitons.

\section{Exciton relaxation}\label{app:relax}

\begin{figure}[h]
 \begin{center}
  \includegraphics[width=0.8\textwidth]{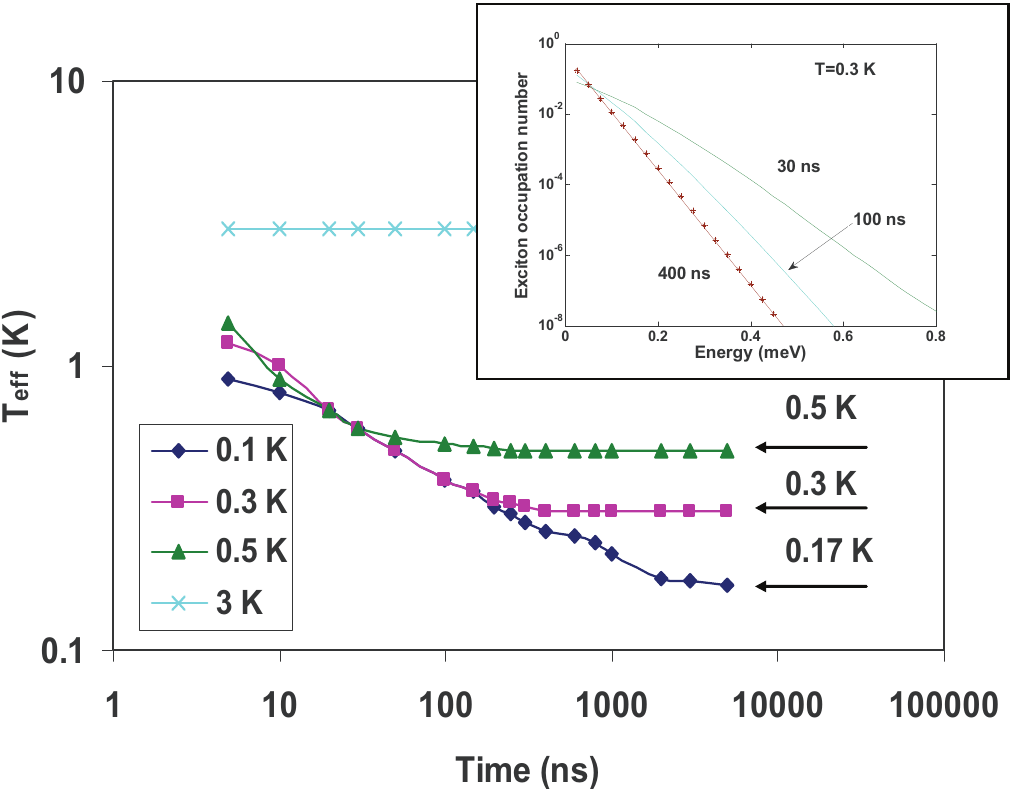}
 \end{center}
 \caption{Cooling behaviour of laser excited paraexcitons for different bath temperatures. Plotted are the effective temperatures of the distribution function obtained by fitting with a Bose distribution. The cooling curve for a lattice temperature of 0.05~K is identical to that of 0.1~K. The inset shows as an example the distribution for a lattice temperature of 0.3~K, an exciton density of $n=1\cdot 10^{15}\,\rm cm^{-3}$ at $t=100$ ns from which a temperature of $T_{\rm eff}=0.33$~K is derived.}
 \label{fig:relaxation0}
\end{figure}

A central problem in the dynamics of excitons at milli-Kelvin temperatures is the relaxation and thermalisation by a contact to a bath of thermal phonons, as the acoustical modes will freeze out \cite{brandt2010}. Therefore, we have simulated the relaxation of hot, laser excited excitons by assuming interactions with longitudinal acoustic phonons and Auger-like two-particle decay in a potential trap by integrating the Boltzmann equation \cite{som2012}. Here we present additional results for the homogeneous situation including elastic exciton-exciton scattering, where the model follows closely that described in \cite{ell1998,snoke1991,ohara2000}. The resulting system of differential equations was integrated using as initial distribution a Gaussian of width 0.1 meV centred at $e_{\rm L}=5$~meV. The initial exciton density was assumed to be $n_{\rm 0} = 1\cdot 10^{15}\,\rm cm^{-3}$. For the elastic exciton-exciton scattering cross section $\sigma=50\,\rm nm^2$ \cite{ohara2000} was taken. 
The main results are shown in figure \ref{fig:relaxation0}, where the effective temperature of the exciton gas obtained by fitting the distribution function by a Bose distribution (see inset) is plotted for different lattice temperatures as a function of time. 

\section{Thermal behaviour of the dilution refrigerator and the sample}\label{app:thermal}

%\subsection*{Description of the dilution system in the cryostat}

\noindent
The $\rm^{3}He/^{4}He$ dilution cryostat uses a mixture of $\rm^{3}He$ and $\rm^{4}He$ for the cooling process. The coldest region is inside the mixing chamber, consisting of a $\rm^{3}He$-rich ({100}{\%} $\rm^{3}He$) and a $\rm^{3}He$-poor (6.4{\%} $\rm^{3}He$) phase separated by a phase boundary, the basic thermodynamics of which is well known \cite{tieftemp}. In the following analysis, we assume that the sample is immersed inside the dilute  $\rm^{3}He/^{4}He$ mixture and that the resistor measuring the bath temperature is placed in between the sample and the phase boundary in the mixing chamber.

\begin{figure}[h]
\begin{center}
\includegraphics[width=10cm]{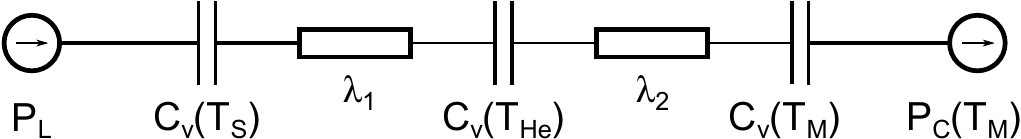}
\caption{Thermal network describing sample and mixing chamber in the $\rm^{3}He/^{4}He$ dilution cryostat. For explanation, see text.}
\label{circuit}
\end{center}
\end{figure}

\begin{figure}[h]
\begin{center}
\includegraphics[width=10cm]{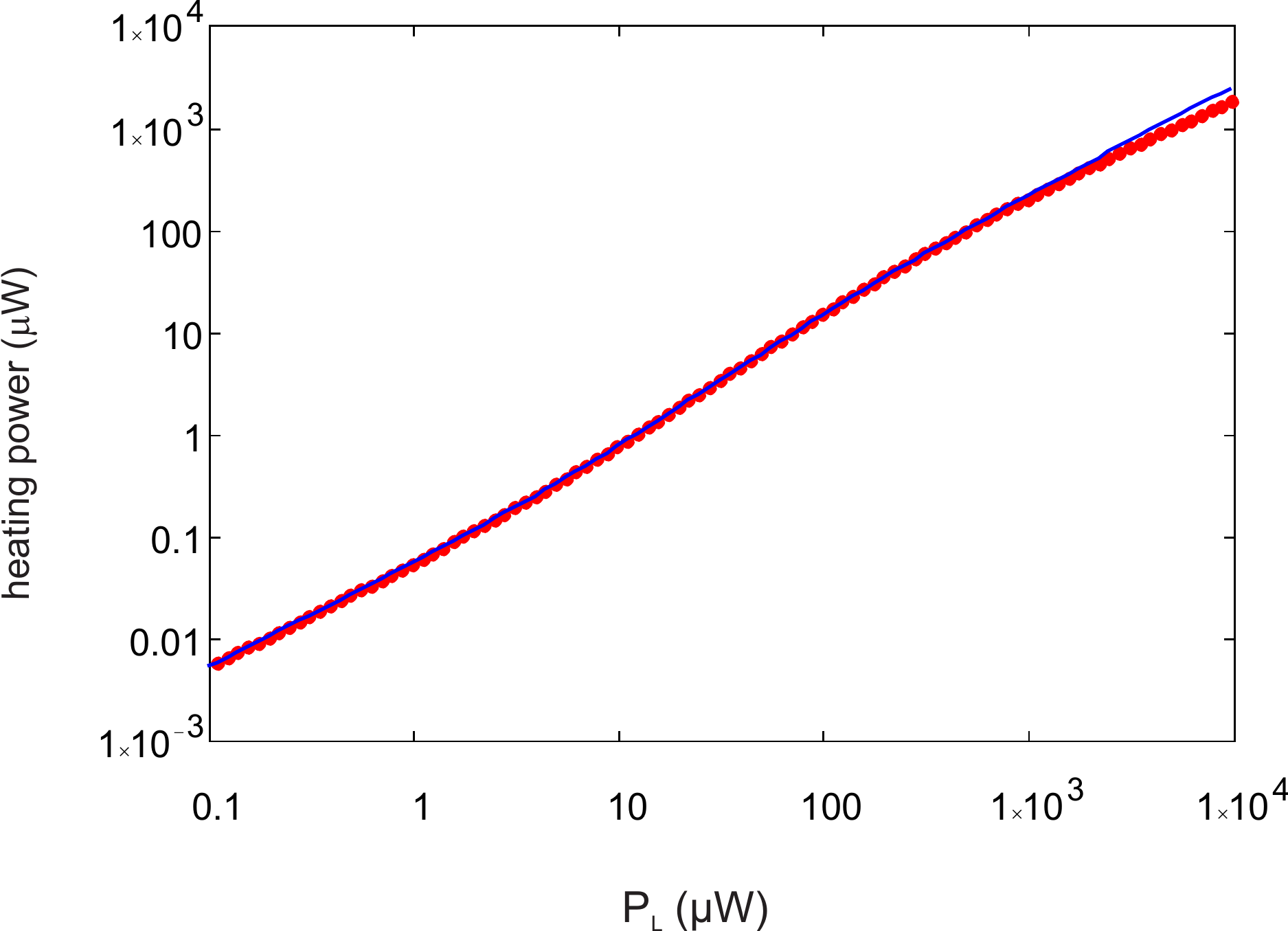}
\caption{Comparison of the amount of heating as a function of laser power. The full red dots are obtained from the rate model of section \ref{sec:ratemodel}, the full line is the fit according to equation \ref{eq:heatfit}.}
\label{fig:heating}
\end{center}
\end{figure}

To derive a connection between sample temperature and measured bath temperature we have to consider the whole system as a thermal network as shown in  figure \ref{circuit}. The incoming laser beam with the power $P_{\rm L}$ hits the sample which is at a temperature $T_{\rm S}$ and has a heat capacity $C_{\rm v}(T_{\rm S})$. It is surrounded by the dilute $\rm^{3}He/^{4}He$ mixture (temperature $T_{\rm He}$), which itself is connected to the mixing chamber at a temperature $T_{\rm M}$ with heat capacities $C_{\rm v}(T_{\rm He})$ and $C_{\rm v}(T_{\rm M})$, respectively. The mixing chamber is cooled with a power $P_{\rm C}(T_{\rm M})$ depending on its temperature. The heat conduction between the different compartments is represented by the heat conductivities $\lambda_1$ and $\lambda_2$. Applying the continuity equation for the energy flow, the network is described by the following system of equations:

 \begin{eqnarray}
C_{\rm v}(T_{\rm S})\cdot\frac{dT_{\rm S}}{dt}&=& F_{\rm heat}(P_{\rm L}) - \lambda_1\cdot(T_{\rm S}-T_{\rm He})\,,\\
\label{diff1}
C_{\rm v}(T_{\rm He})\cdot\frac{dT_{\rm He}}{dt}&=&\lambda_1\cdot(T_{\rm S}-T_{\rm He})-\lambda_2\cdot(T_{\rm He}-T_{\rm M})\,,\\
\label{diff2}
C_{\rm v}(T_{\rm M})\cdot\frac{dT_{\rm M}}{dt}&=&\lambda_2\cdot(T_{\rm He}-T_{\rm M})-P_{\rm C}(T_{\rm M})\,.
\label{diff3}
\end{eqnarray}
Here $F_{\rm heat}$ gives the connection between the laser input power and the heat generated in the sample due to the excitonic relaxation and decay processes (compare section \ref{sec:ratemodel}). For the parameters given in table 1, the results are shown in figure \ref{fig:heating}. We can approximate the dependence by a function of the form
\be \label{eq:heatfit}
F_{\rm heat}= \alpha P_{\rm L} + \beta\left[ P_{\rm XX}\left(\sqrt{1+P_{\rm L}/P_{\rm XX}}-1 \right)\right]^2
\ee
with the fit parameters $\alpha=0.045, \beta=0.014\,\micro$m$^{-1}$, and $P_{\rm XX}=16.67\,\micro$W. 

\subsection*{Heat capacities}
Regarding the heat capacity of the {\it sample}, it is necessary to take into consideration that not only the crystal is heated by the laser, but, due to the good thermal contact supplied by the pressure on the sample, also the sample holder, which consisted of pure titanium. Therefore, we assumed the temperatures of sample and sample holder to be the same. The variation of the heat capacity of the sample $\rm Cu_{2}O$ with temperature is well approximated by the Debye model as a $T^3$ temperature dependence. A fit of measured data \cite{Cu2O} by
\be
C_{\rm v_{\rm Cu_{2}O}}=A\cdot T^{B},
\label{CvCu2O}
\ee
results in the parameters $A=8.06\pm0.05\, {\rm J}/{\left(\rm mol\,K^4\right)}$ and $B=3.0\pm0.05$.
For the specimen holder, we consider only the superconducting state below the critical temperature ($T_{\rm c_{\rm Ti}}=0.4\, \rm{K}$ \cite{Ti}), restricting the following analysis to the interesting case $T_{\rm S}<0.4 \rm K$. Here, the heat capacity is composed of an electron and a phonon part, $C_{\rm v_{\rm Ti}}=C_{\rm v_e}+C_{\rm v_{\rm ph}}$. While the phonon part is negligible, the electron part is given by
\be
C_{\rm v_{\rm Ti}}=C_{\rm v_e}= C_{\rm Ti}\, \gamma_{\rm Ti}\, T_{\rm c}\, \exp\left(\frac{-1.5 T_{\rm c}}{T}\right) ,
\ee
with $C_{\rm Ti}=9.17 $ and the Sommerfeld constant $\gamma_{\rm Ti}=3.3\,\rm {mJ}/({mol \cdot K^2})$ \cite{Ti}.

%\subsection*{Heat capacity of the helium mixture}
\noindent
Experimental data for the {\it heat capacity of the helium mixture} \cite{He} for different concentrations of $\rm^{3}He$ in $\rm^{4}He$ were interpolated to a concentration of $x_{\rm^{3}He}=6.4{\%}$ and fitted resulting in a dependence
\be
C_{\rm v}(T_{\rm He, M})=D\cdot\left(\frac{T_{{\rm He}, M}}{\rm K}\right)^{E}+F ,
\label{Cv(THe)}
\ee
with $D={0.417}\,{\rm J}/{(\rm mol\,K)}$, $E={6.117}$ and $F={0.7}\,{\rm J}/{(\rm mol\,K)}$.

\subsection*{Cooling power}
For the description of the cooling power $P_{\rm C}$ of the dilution process, where the $\rm^{3}He$-atoms flow from the $\rm^{3}He$-rich (concentrated) in the -poor (dilute) phase, we consider the enthalpy of the system. By assuming that pure $\rm^{3}He$ flows through the heat exchangers into the mixing chamber with the dilute phase, the cooling power is given in accordance with \cite{tieftemp} by:
\bea
P_{\rm C}&=&\dot{n}_3\cdot[H_{\rm d}(T)-H_{\rm c}(T)]\\
&=&\dot{n}_3\cdot[95\cdot T_{\rm M}^{2}-11\cdot T_{\rm W}^{2}],
\label{PC}
\eea
with the molar flow rate $\dot{n}_3$, the temperature in the mixing chamber $T_{\rm M}$ and the temperature behind the heat exchanger $T_W$.
To obtain the unknown parameters of equation (\ref{PC}), we take as calibration points the lowest temperature reached without any heat load $T_{\rm M}=20\,\rm mK$ and $P_{\rm C}=100\,\micro$W at $T_{\rm M}=100\,\rm mK$. From these data we obtain $\dot{n}_3={1.25}\cdot 10^{-4}\,\rm{mol}/{s}$ and $T_W={0.115}\,\rm{K}$. From this cooling power we have to subtract the heat load $P_{\rm 0}$ due to the cryostat windows, which lead to a minimum temperature of $T_{\rm M}=38\,\rm mK$ corresponding to a power of $P_{\rm 0}=13.4\,\micro\rm W$.
\noindent

\subsection*{Determination of the temperatures in the cryostat system}
To obtain the temperatures of the different parts of the thermal network from the system of coupled differential equations, it is necessary to specify the amount of substances in the system.

\begin{figure}[h]
\begin{center}
\includegraphics[width=0.9\textwidth]{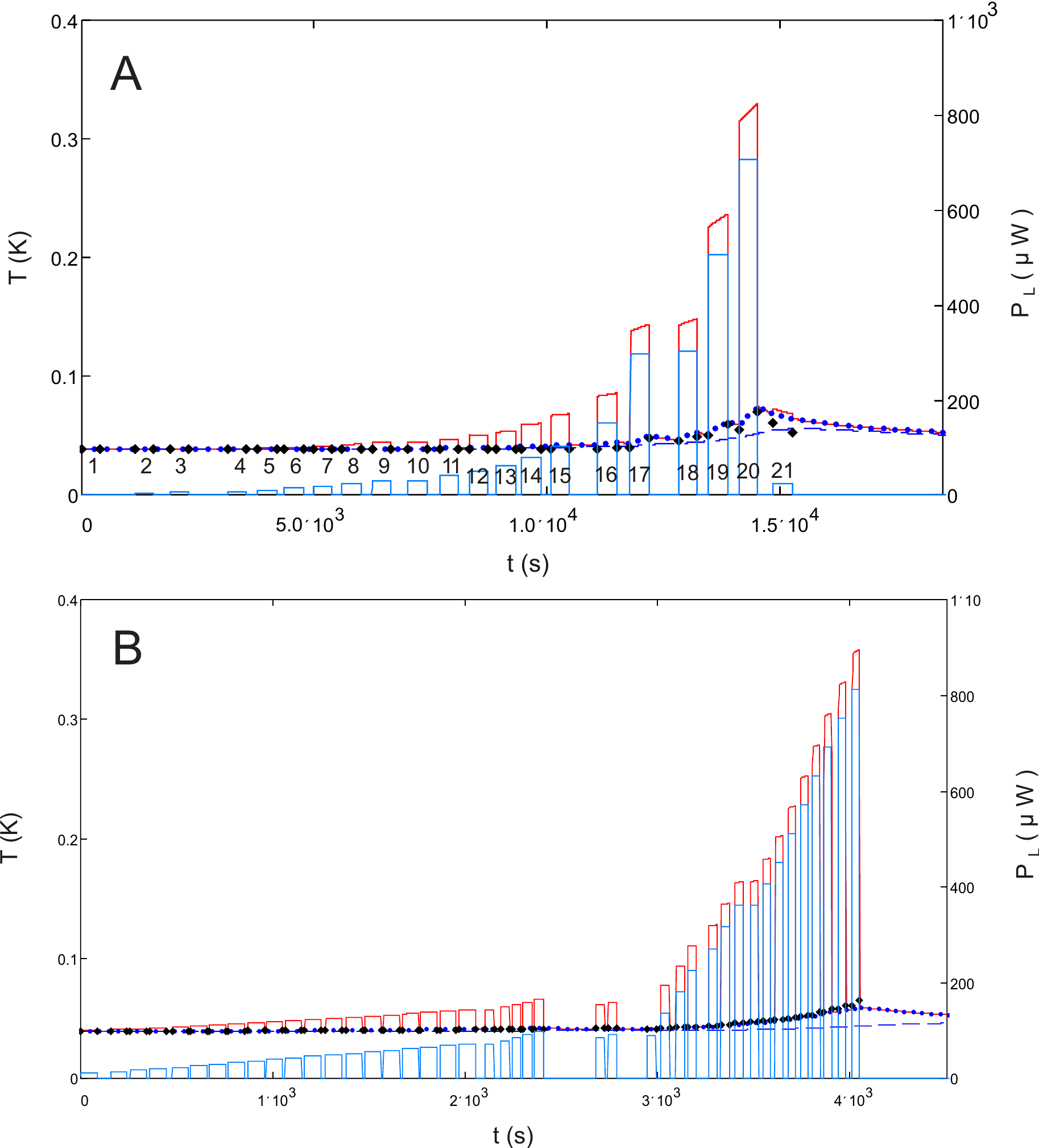}
\caption{The numerical solution of the system of differential equations expresses the variation of the different temperatures (cf.\ equations (\ref{diff1})-(\ref{diff3})) with the measurement time for $\lambda_1=6\cdot10^{-4}\,\rm W/K$ and $\lambda_2=27\cdot10^{-4}\,\rm W/K$. The solid red framed bars represent the temperature of the sample $T_{\rm S}$, the blue dotted curve shows the temperature of the surrounding helium bath $T_{\rm He}$ and the blue dashed line describes the temperature in the mixing chamber $T_{\rm M}$. The solid blue framed bars are the representation of the measured incoming laser power $P_{\rm L}$. The measured helium temperatures are shown as black diamonds.}
\label{fig:Temp}
\end{center}
\end{figure}
\noindent
The $\rm Cu_{2}O$ specimen is a cube with an edge length of 3 mm, a molar mass of $M={143.09}\,\rm {g}/{mol}$ and a density of $\rho=6.48\,\rm{g}/{cm^{3}}$. Hence, the amount of substance inside the cube is $n_{\rm Cu_{2}O}={1.398}\cdot10^{-3}\,\rm mol$. From the measured weight of the titanium sample holder of $m=26\,\rm g$ and a molar mass of $M={47.9}\,\rm {g}/{mol}$, we obtain $n_{\rm Ti}={0.5}\,\rm mol$.
The amount of substance in the helium mixture can only be estimated. In our system, a 15 \% helium $\rm ^3He/^4He$ mixture of approximately $V=400\,\rm l$ circulates under a pressure of $p={0.8}\,\rm bar$. That leads to an amount of helium mixture of $n_{\rm ^{3}He/^{4}He}={14.3}\,\rm mol$ from which only about two thirds are in the mixing chamber.
For our calculations, we assume that 1 mol is in the bath and {2.75} mol is in the mixing chamber.
%\subsubsection*{The temperatures}
Then the only quantities which have still to be specified are the heat conductivities $\lambda_{1,2}$. This was done by adjusting the solutions of the coupled differential equations to different measurements until all data could be described by the same set of parameters. As examples we show in Figure \ref{fig:Temp} the results for the series of measurements from Figure \ref{fig:powerspectra} and \ref{fig:power_temp}.

The calculations were done with $\lambda_1=6\cdot10^{-4}\,\rm {W}/{K}$ and $\lambda_2=27\cdot10^{-4}\,\rm {W}/{K}$ and show an almost quantitative agreement with the experimental data, despite very large changes in laser power (solid blue framed bars) during the measurements, which indicates the correctness of our model.

\section*{References}

\bibliographystyle{unsrt}
\bibliography{Biblio_DYN_in_trap}

\end{document}